\documentclass[secnumarabic,aps,prb,reprint]{revtex4-2}
\usepackage[english]{babel}
\usepackage{graphicx}
\usepackage{color}
\usepackage{amsmath}
\usepackage{tabularx}
\usepackage[dvipsnames]{xcolor}

\begin{document}

	\title{Effects of spin-orbit interaction and electron correlations in strontium titanate}
	\author{Sergei Urazhdin}
	\affiliation{Department of Physics, Emory University, Atlanta, GA, USA.}
	\author{Ekram Towsif}
	\affiliation{Department of Physics, Emory University, Atlanta, GA, USA.}
	\author{Alexander Mitrofanov}
	\affiliation{Department of Physics, Emory University, Atlanta, GA, USA.}
	
\begin{abstract}
We show that the Bloch states in the conduction band of SrTiO$_3$ arise from the interplay between highly anisotropic hopping in sub-bands derived from the Ti $t_{2g}$ orbitals and spin-orbit coupling that mixes these orbitals. Because of the nearly flat-band characteristics for one of the principal axes, at sufficiently high doping these Bloch states become unstable with respect to electron interactions, resulting in Mott-like singlet correlations. These findings may be relevant to the anomalous electronic properties of SrTiO$_3$, including its unusual superconductivity.
\end{abstract}

\maketitle

\section{Introduction}\label{sec:intro}

Strontium titanate, SrTiO$_3$ (STO), exhibits unique and puzzling electronic and structural properties, which have motivated its extensive studies over the last 50 years~\cite{RevModPhys.60.585, condmat5040060, GASTIASORO2020168107, doi:10.1146/annurev-conmatphys-031218-013144}. The dielectric constant of STO is anomalously large and almost diverges at low temperatures without the onset of ferroelectricity in a manner consistent with quantum paraelectricity~\cite{PhysRevB.19.3593}. Strain or interfacial effects in thin films can stabilize ferroelectricity~\cite{PhysRevB.13.271, BURKE1971191}.

Electron-doped STO also exhibits superconductivity (sc) at record-low carrier concentrations $n\gtrsim 3\times10^{17}$~cm$^{-3}$, corresponding to the Fermi energy of less than $2$~meV~\cite{PhysRevB.19.3593, PhysRevLett.112.207002}. Experiments suggest s-wave symmetry of the sc order parameter~\cite{PhysRevB.92.174504, PhysRevB.90.140508}. Furthermore, the large lattice fluctuations associated with quantum paraelectricity in STO are suggestive of the conventional phonon mechanism of sc. However, sc in STO cannot be explained by the usual Migdal-Eliashberg extension of the Bardeen-Cooper-Schrieffer (BCS) theory relying on the electron attraction mediated by the retarded lattice response~\cite{tinkham2004introduction,Marsiglio2020}, since the Fermi energy in STO is comparable to the energy of phonons~\cite{doi:10.1143/JPSJ.49.1267, Kirzhnits1973}.

The dome-like dependence of the critical temperature $T_c$ on doping is similar to that of high-temperature superconductors (HTSCs), albeit with a much smaller maximum $T_c=0.4$~K~\cite{PhysRevLett.12.474, PhysRev.163.380}. Furthermore, tunneling measurements indicate multi-band sc, similar to some unconventional superconductors, such as ruthenates and pnictides~\cite{PhysRevLett.45.1352}. A variety of the proposed mechanisms include long-range electron-phonon interaction~\cite{osti_5140639}, soft bosonic modes~\cite{PhysRevLett.115.247002}, intervalley phonons~\cite{PhysRevLett.21.16}, and quantum paraelectric fluctuations~\cite{Rischau2017, Rowley2014}, but the mechanism of sc in STO is still debated. In particular, it remains contentious whether the ferroelectric distortions enhance sc in STO~\cite{Ahadi2019} or suppress it~\cite{PhysRevLett.115.247002,PhysRevResearch.4.013019}.

Here, we present a tight-binding analysis of the Bloch states in the conduction band, which may shed light on the puzzling electronic properties of STO and its heterostructures. In the next section, we show that in the limit of negligible spin-orbit coupling (SOC), the sub-bands derived from the three $t_{2g}$ orbitals of Ti are highly anisotropic. In Section~\ref{sec:SOC}, we show that near the bottom of the conduction band, these states are mixed by SOC into a Kramers doublet with the total moment $j=5/2$. In Section~\ref{sec:interactions}, we utilize the Hubbard model to show that at sufficiently high carrier density, the large sub-band anisotropy results in the instability of the Bloch states with respect to Mott-like singlet correlations.  Finally,  in Sections \ref{sec:sc} and \ref{sec:discussion}, we discuss the relation of these results to sc and other anomalous electronic properties of STO.

\section{Conduction band without SOC}\label{sec:1p}

 In this section, we show that in the limit of negligible SOC, the conduction band structure of STO is determined by highly anisotropic sub-band hopping. While the band structure of STO has been extensively studied~\cite{PhysRevB.6.4740,Guo2003,Piskunov2004,PhysRevB.84.205111, PhysRevB.94.035111}, we are not aware of prior studies of this property revealed by our analysis. We argue below that it is important for understanding the effects of lattice distortions, confinement at interfaces, and the role of SOC.

\begin{figure}
	\centering
	\includegraphics[width=1.0\columnwidth]{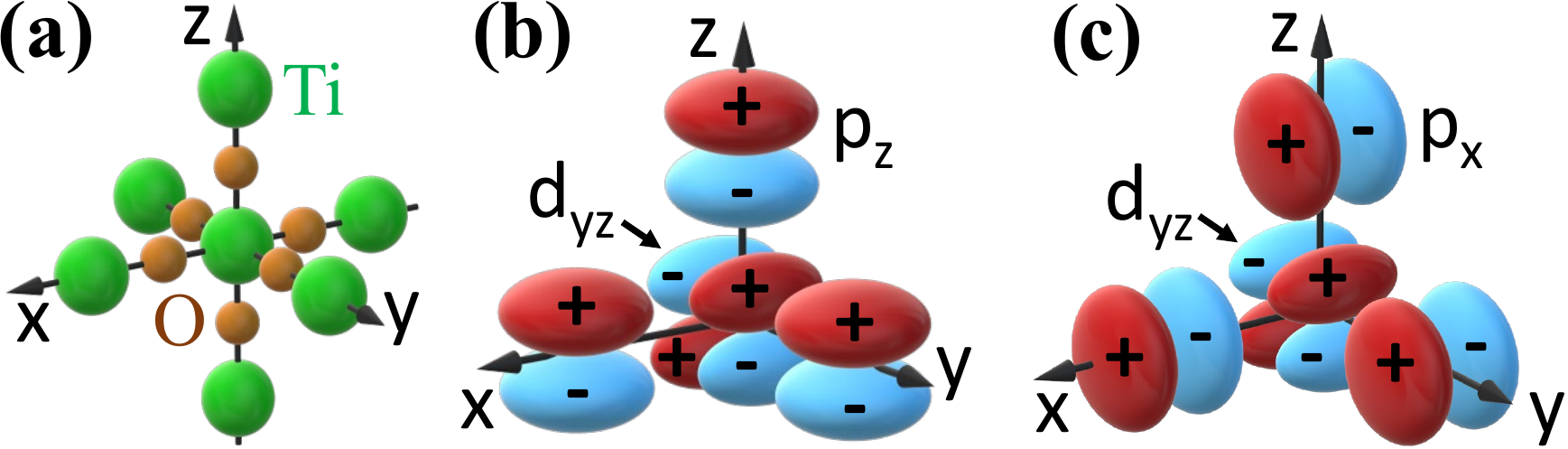}
	\vspace{-5pt}
	\caption{\label{fig:orbitals} (a) The structural motif dominating the conduction band in STO consists of Ti-O-Ti chains along the principal axes, with octahedral coordination of Ti by oxygen atoms. (b), (c) Schematics of the $d_{yz}$ orbital of Ti and $p_z$ (b), $p_x$ (c) orbitals of nearest-neighbor oxygens on three principal axes. The signs of the orbital lobes are labeled.}
\end{figure}

At room temperature, STO has a cubic perovskite structure, with octahedral coordination of Ti formed by six nearest-neighbor oxygens aligned with the principal crystal axes, Fig.~\ref{fig:orbitals}(a). Antiferrodistortive rotations of the ${\mathrm{TiO}}_{6}$ octahedra below $105$~K result in minor distortions of the octahedral environment of Ti~\cite{PhysRev.177.858}. These distortions slightly reduce the orbital selectivity of hopping discussed below, which does not qualitatively change our findings. On the other hand, uniaxial strain in thin films or ferroelectric distortions due to the quantum paraelectric fluctuations at cryogenic temperatures may have a significant effect on the conduction band structure, as discussed in Section~\ref{sec:interactions}. 

The conduction band of STO is mainly derived from hopping between the $t_{2g}$ orbitals of Ti and the p-orbitals of its nearest-neighbor oxygen atoms. This hopping can be characterized by the matrix elements $t^{i}_{m,m'}=\langle d_m|V'|p^i_{m'}\rangle$~\cite{pavarini2016quantum}. Here, the index $i$ enumerates the nearest-neighbor oxygen atoms, $d_m$ is one of the three $t_{2g}$ orbital wavefunctions of Ti forming a pseudo-vector $(d_{yz}$,$d_{xz}$,$d_{xy})=(d_1,d_2,d_3)$, the index $m'$ enumerates oxygen's p-orbitals, and $V'$ is the perturbation of the atomic potential resulting in orbital hybridization. We consider the additional effects of O-O hopping separately at the end of this section, and SOC in the next section. 

We now show that Ti-O hopping is described by a single orbitally selective matrix element. Analysis of the Koster-Slater parameters yields the same result~\cite{Harrison}, but does not reveal the underlying symmetries. Consider the $d_{yz}$ orbital of Ti and the $p_z$ orbital of the neighboring oxygen on the x-axis, Fig.~\ref{fig:orbitals}(b). The dependence of the wavefunction $d_{yz}$ on the axial angle $\theta_x$ for rotations around the x-axis is $\sin(2\theta_x)$. Meanwhile, for the $p_z$ orbital of oxygen on the x-axis, this dependence is $\sin(\theta_x)$. In the cubic phase, the potential $V'$ is axially symmetric. In the cylindrical coordinate system $(x,\rho_x,\theta_x)$ aligned with the x-axis, the corresponding matrix element is $t^{1}_{1,3}=\int f(x,\rho_x)\sin(\theta_x)\sin(2\theta_x)d\theta_x=0$, where $f(x,\rho_x)$ is a function of radial and axial coordinates. 

The hopping amplitude $t^{1}_{1,3}$ vanishes by symmetry, since each lobe of the $p_z$ orbital has the same overlap with both the positive and the negative lobes of the $d_{yz}$ orbital,  Fig.~\ref{fig:orbitals}(b). 
Symmetry also prohibits hopping between the $d_{yz}$ orbital and the $p_z$ orbital of the oxygen on the $z$-axis. The only symmetry-allowed matrix element involving the $d_{yz}$ orbital, $t^{2}_{1,3}$, describes hopping to the $p_z$ orbital of nearest-neighbor oxygen atoms on the y-axis, since both orbitals are described by the same dependence $\cos(\theta_y)$ on the rotation angle around the y-axis. By the cubic symmetry, the only finite matrix element involving hopping between the $d_{yz}$ orbital of Ti and the $p_y$ orbitals of oxygen atoms is $t^{3}_{1,2}=t^{2}_{1,3}$, which corresponds to hopping along the z-axis. Symmetry also prohibits  hopping between the $d_{yz}$ and $p_x$ orbitals in any direction, Fig.~\ref{fig:orbitals}(c).

We conclude that an electron in the $d_{yz}$ orbital of Ti can hop only onto the $p_z$ orbital of the two nearest-neighbor oxygen atoms on the y-axis or the $p_y$ orbital of the two nearest-neighbor oxygen atoms on the z-axis. By the same symmetry arguments, it can then hop from oxygen only onto the $d_{yz}$ orbital of the nearest-neighbor Ti along the corresponding Ti-O-Ti chain [see Fig.~\ref{fig:orbitals}(b)]. Thus, electrons initially in the $d_{yz}$ orbital propagate only in the $yz$-plane while retaining their orbital state.

In the considered approximation, oxygen atoms merely mediate orbitally-selective hopping between Ti atoms. We can then consider only the state projections on the Ti $t_{2g}$ orbitals. Orbital state-preserving hopping between the $d_{yz}$ orbital of Ti and its four nearest Ti neighbors in the $yz$-plane is described by a single hopping parameter $t$, which is negative since $d_{yz}$ is antisymmetric with respect to both the y- and the z-axes [see Fig.~\ref{fig:orbitals}(b)]. By symmetry, hopping on the $d_{xz}$ and $d_{xy}$ orbitals occurs only within the $xz$- and $xy$-planes, respectively. The corresponding Hubbard Hamiltonian is

\begin{equation}\label{eq:H_hop}
\begin{split}
\hat{H}_{hop}=&\epsilon_0\sum_{\vec{n},m,s}\hat{n}_{\vec{n},m,s}\\
	+&t\sum_{\vec{n},\vec{l},m,s} {(1-\delta_{l,m})\hat{c}^+_{\vec{n}+\vec{l},m,s}\hat{c}_{\vec{n},m,s}},
\end{split}
\end{equation}
where $\epsilon_0$ is the level energy, $\hat{c}^+_{\vec{n},m,s}$ is the electron creation operator on site $\vec{n}$ in orbital $d_m$ with projection $s=\pm1/2$ of spin on the z-axis, $\hat{n}_{\vec{n},m,s}=\hat{c}^+_{\vec{n},m,s}\hat{c}_{\vec{n},m,s}$, 
$\vec{l}$ is a unit vector in one of the two directions along the $l^{th}$ principal axis, and $\delta_{l,m}$ is the Kronecker symbol.

The single-particle eigenstates derived from the orbitals $d_m$ are Bloch waves
\begin{equation}\label{eq:H_hop_sol}
	\psi_{\vec{k},m,s}=\frac{1}{\sqrt{N}}\sum_{\vec{n}} e^{ia\vec{k}\vec{n}}\hat{c}^+_{\vec{n},m,s}|0\rangle=\hat{c}^+_{\vec{k},m,s}|0\rangle,
\end{equation}
where $a$ is the lattice constant, and $N$ is the total number of sites. Their dispersion is
\begin{equation}\label{eq:dispersion_noSOC}
E_m(\vec{k})=\epsilon_0+2t\sum_{m'}(1-\delta_{m,m'})\cos(k_{m'}a).
\end{equation}
We choose $\epsilon_0=-4t$ so that $\hat{H}_{hop}$ is the kinetic energy.

\begin{figure}
	\centering
	\includegraphics[width=0.9\columnwidth]{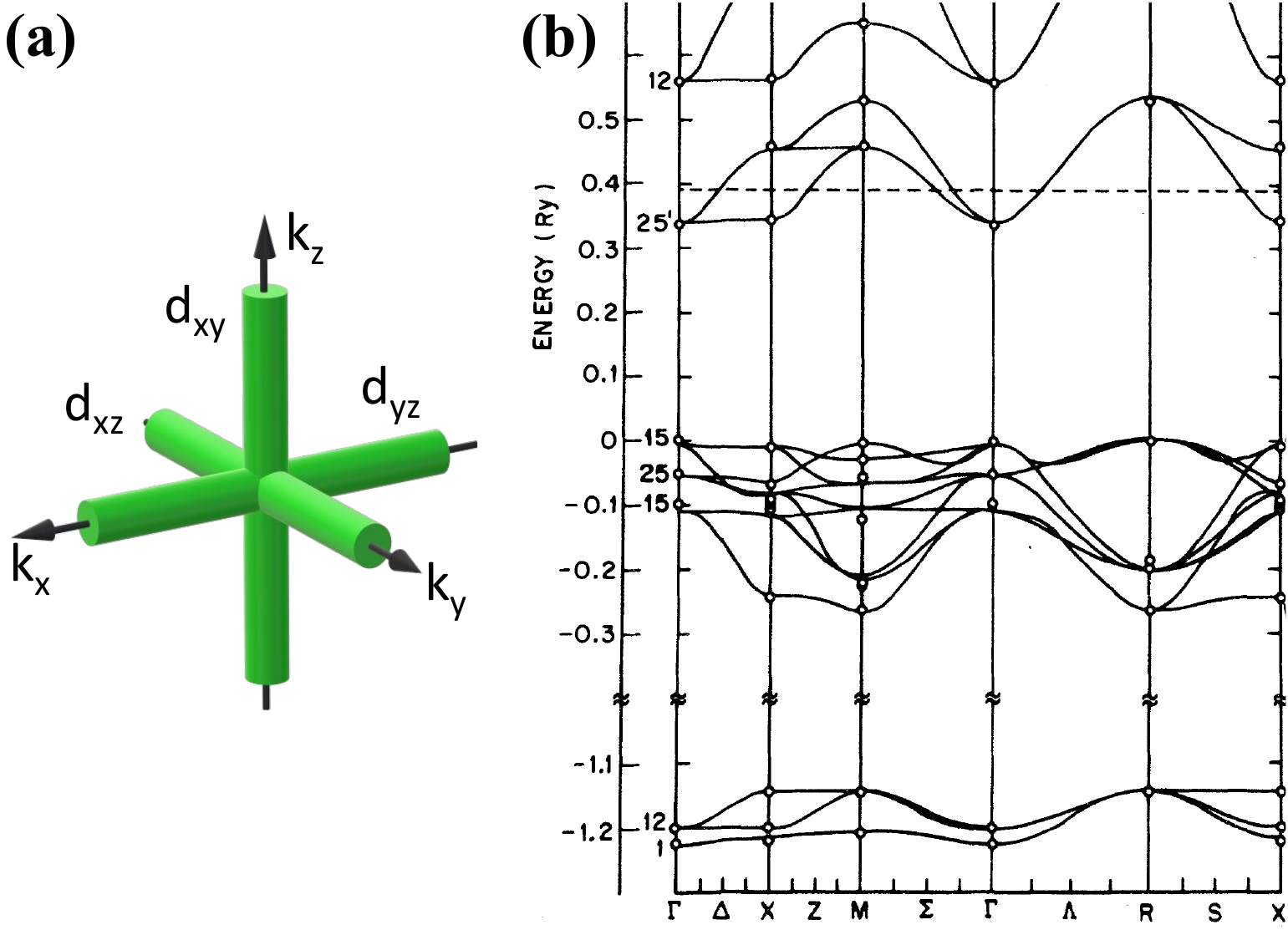}
	\caption{\label{fig:bands} (a) In the limit of negligible SOC, the Fermi surfaces of the sub-bands derived from the orbitals $d_m$ are cylinders spanning the Brillouin zone and aligned with the principal axes. (b) LCAO band structure reproduced from Ref.~\cite{PhysRevB.6.4718}.}
\end{figure}

In the discussed approximation, the $m^{th}$ sub-band is non-dispersive in the $m^{th}$ direction and is parabolic at small $k$ in the other two directions. The resulting Fermi surface consists of three orbitally-selective cylindrical sub-surfaces aligned with the reciprocal axes that span the Brillouin zone, Fig.~\ref{fig:bands}(a).

\textbf{Effects of oxygen-oxygen hopping.}  We now address the possibility that other hopping contributions, and especially nearest-neighbor O-O hopping whose amplitude is smaller but not negligible compared to Ti-O hopping, may affect the sub-band anisotropy identified above. Figure~\ref{fig:bands}(b) reproduces the band structure obtained in Ref.~\cite{PhysRevB.6.4718} using the linear combination of molecular orbitals (LCAO) method, which included O-O hopping, along with other hopping contributions among the $14$ orbitals considered in the model. Remarkably, one of the conduction sub-bands is nearly flat in the principal $\Gamma-X$ direction, in good agreement with the above analysis. Meanwhile, the other sub-band is highly dispersive in this direction and is doubly degenerate, as is apparent from its splitting along the $X-Z$ direction. We conclude that the sub-band anisotropy identified above based on the analysis of Ti-O hopping is robust with respect to other hopping contributions.

We now show that O-O hopping does not compromise the orbitally-selective anisotropy of hopping because of the cubic symmetry. O-O hopping is characterized by two matrix elements corresponding to $\sigma$ and $\pi$ bonds, whose amplitudes $t_{\sigma}=0.4$~eV, $t_{\pi}=-0.04$~eV, were determined in LCAO calculations~\cite{PhysRevB.6.4718,Harrison}. The value of $t_{\pi}$ is an order of magnitude smaller than $t_{\sigma}$. Since the oxygen amplitudes in the conduction band are small, O-O hopping via the $\pi$ bonds can be neglected.

We enumerate the oxygen sites coordinating the $\vec{n}^{th}$ Ti atom with indices $\vec{n}+\vec{l}/2$, where $\vec{l}$ is a unit vector in one of the principal directions. The O-O hopping Hamiltonian via $\sigma$ bonds can be written as 
\begin{equation}\label{eq:H_OO}
	\hat{H}_{O-O}=t_{\sigma}\sum_{\vec{l}'\ne\vec{l},s}\hat{a}^+_{\vec{n}+\vec{l}/2,\vec{u},s}\hat{a}_{\vec{n}+\vec{l}'/2,\vec{u},s},
\end{equation}
where the operator $\hat{a}^+_{\vec{n}+\vec{l}/2,\vec{u},s}$ creates an electron with spin $s$ on the oxygen site $\vec{n}+\vec{l}/2$ in the state $p_{z'}$, with the local axis $z'$ pointing in the direction of the unit vector $\vec{u}=(\vec{l}-\vec{l}')/\sqrt{2}$ connecting the two oxygen atoms. In the principal axes coordinates, this Hamiltonian is

\begin{equation}\label{eq:H_OO2}
\begin{split}
	\hat{H}_{O-O}=t_{\sigma}\sum_{\vec{l}'\ne\vec{l},m,m',s}\hat{a}^+_{\vec{n}+\vec{l}/2,m,s}\hat{a}_{\vec{n}+\vec{l}'/2,m',s}\\
	\times\left[ u_mu_{m'}	-\delta_{mm'}\frac{(1-|l_m|)(1-|l'_m|)}{8}\right]
	,
\end{split}
\end{equation}

where the operator $\hat{a}^+_{\vec{n}+\vec{l}/2,m,s}$ creates an electron with spin $s$ on the oxygen site $\vec{n}+\vec{l}/2$ in the state $p_m$. In this equation, the terms with $m'\ne m$ mix different p-orbitals, while the terms with $m'=m$ conserve the orbital state. 

To show that the Hamiltonian Eq.~(\ref{eq:H_OO2}) does not violate the hopping anisotropy identified above, we consider a specific case of hopping in the z-direction for the sub-band $d_3$ derived from the Ti $d_{xy}$ orbital. Such hopping is not allowed by Ti-O hopping alone. It is easy to see that for hopping along the principal axes, the orbital-mixing contribution vanishes because the hopping amplitude changes sign upon the reversal of $u_m$ or $u_{m'}$.

\begin{figure}
	\centering
	\includegraphics[width=0.45\columnwidth]{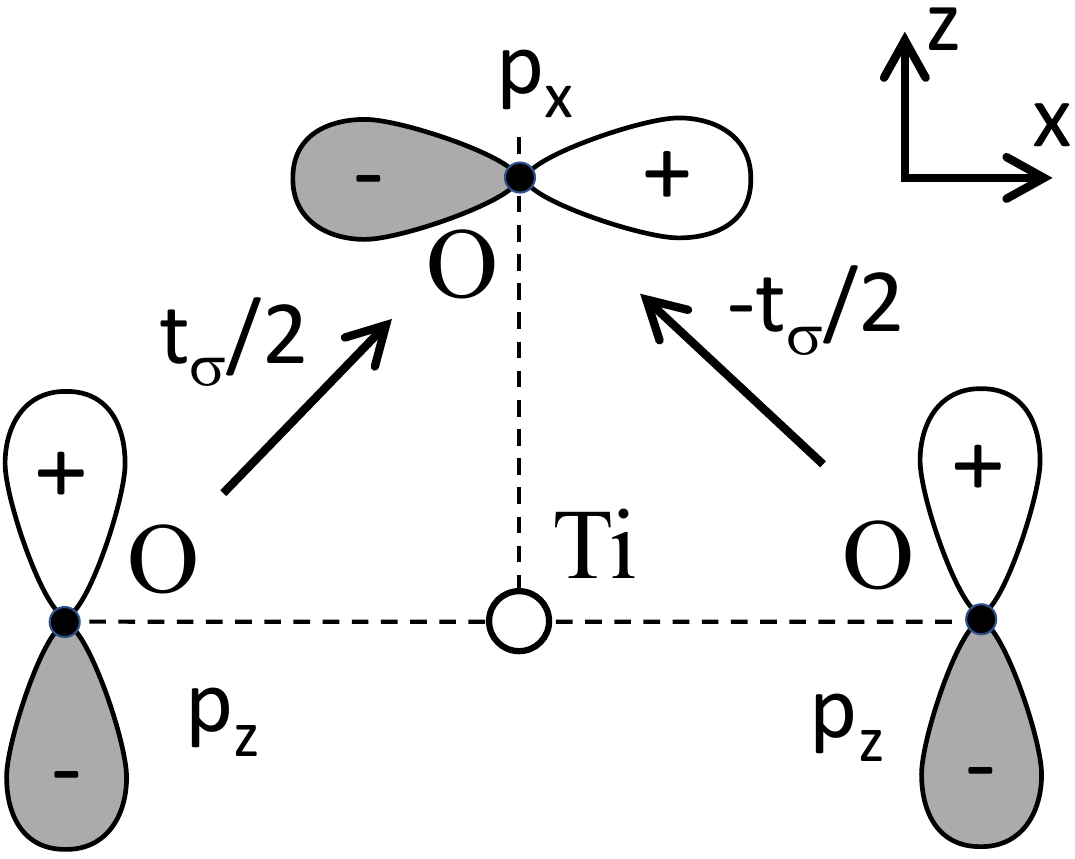}
	\caption{\label{fig:mixing} Illustration of destructive interference between two diagonal O-O hopping directions suppressing orbital mixing for the wavevector along the z-axis.}
\end{figure}

The underlying mechanism is illustrated in Fig.~\ref{fig:mixing} for hopping in the x-z plane between the $p_z$ and $p_x$ orbitals. Hopping from the $p_z$ orbital to the $p_x$ orbital in the positive-z, positive-x direction is characterized by the amplitude $t_\sigma/2$, while hopping in the positive-z, negative-x direction is characterized by the amplitude $-t_\sigma/2$. The amplitudes are opposite because the orbital $p_x$ is antisymmetric with respect to x-axis inversion. Destructive interference between the two contributions prevents orbital mixing. We note that if the wavevector has a finite x-component, a complete cancellation does not occur, and the hybridization between the oxygen orbitals shown in Fig.~\ref{fig:mixing} provides a non-negligible contribution to hopping. This results in the renormalization of the dispersion Eq.~(\ref{eq:dispersion_noSOC}) but does not affect our conclusions.

We now consider the remaining O-O hopping contributions, which conserve orbital moment. The $d_3$ sub-band formed by Ti-O bonding hybridizes the Ti $d_{xy}$ orbitals only with the $p_x$ and $p_y$ orbitals of oxygen, and therefore the terms with $m=m'=3$ in Eq.~(\ref{eq:H_OO2}) do not have finite matrix elements with this sub-band. 

The remaining terms describe hopping on the oxygen's $p_x$ and $p_y$ orbitals. Their contribution to the $d_3$ sub-band dispersion also vanishes, because the hopping amplitudes $t_\sigma/2$ for $m'=m$ in the first term in the square brackets in Eq.~(\ref{eq:H_OO2}) are opposite to the amplitudes in the second term, resulting in destructive interference between hopping in the xz- and yz-planes, whose mechanism is similar to the suppression of inter-orbital hopping illustrated in Fig.~\ref{fig:mixing}. We conclude that by the symmetry, O-O hopping does not compromise the anisotropy of orbitally-selective sub-band hopping identified in our analysis of Ti-O hopping. This conclusion is also supported by the \textit{ab initio} calculations~\cite{PhysRevB.84.205111}, as discussed in the next section.

\section{SOC effects}\label{sec:SOC}

The atomic SOC Hamiltonian is $\hat{H}_{SO}=-\lambda\vec{\hat{L}}\cdot\vec{\hat{s}}$, where $\lambda\approx 18$~meV for Ti~\cite{Dunn1961-ce}, $\vec{\hat{L}}$ and $\vec{\hat{s}}$ are the orbital and spin angular momenta in units of Planck's constant. The projection of this Hamiltonian onto the $t_{2g}$ subspace is
\begin{equation}\label{eq:H_SO}
\hat{H}_{SO}=i\frac{\lambda}{2}\sum_{\vec{k},m_i,s,s'} e_{m_1m_2m_3}\sigma^{m_1}_{ss'}\hat{c}^+_{\vec{k},m_2,s}\hat{c}_{\vec{k},m_3,s'},
\end{equation}
where $e_{m_1m_2m_3}$ is the Levi-Civita symbol and $\sigma^{m}$ is the $m^{th}$ Pauli matrix. 

The anisotropy of hopping does not permit a general analytical solution for $\hat{H}=\hat{H}_{hop}+\hat{H}_{SO}$ at arbitrary $\vec{k}$. Thus, we separately consider the analytically solvable limiting regimes, and use a perturbative approach to interpolate between them. For $\vec{k}$ along $k_z$, the $d_1$ and $d_2$ sub-bands are degenerate and split from the $d_3$ sub-band by $\Delta E\approx ta^2k^2$. According to the perturbation theory, their mixing with the $d_3$ sub-band is negligible at $k\gg a^{-1}\sqrt{\lambda/t}$.

The eigenstates of $\hat{H}_{SO}$ on the subspace of orbitals $d_1$, $d_2$ are 
Bloch waves with the atomic orbital structure $d_{\pm}=(d_1\mp id_2)/\sqrt{2}$ characterized by definite orbital moment projections $M=\pm1$ on the z-axis. SOC splits the $d_1$, $d_2$ band into a two-fold degenerate $j=1/2$ sub-band derived from the spin-orbit coupled atomic states $d_-\uparrow$, $d_+\downarrow$, and a $j=3/2$ sub-band derived from the states $d_+\uparrow$, $d_-\downarrow$, where the up and down arrows denote the spin direction with respect to the z-axis. The two sub-bands are split by $\lambda$.

At $k=0$, the kinetic energy vanishes and $\hat{H}$ can be diagonalized. The ground state (g.s.) is split by SOC into two levels. The first level is a Kramers doublet with atomic spin-orbital structure 
\begin{equation}\label{eq:psi_sigma}
\begin{split}
\psi_+=(\sqrt{2}d_+\uparrow-d_3\downarrow)/\sqrt{3},\\
\psi_-=(\sqrt{2}d_-\downarrow+d_3\uparrow)/\sqrt{3},
\end{split}
\end{equation}
with $j=5/2$ and energy $\epsilon_\pm=-\lambda$. The second level characterized by atomic moment $j=1/2$ and energy $\epsilon_n=\lambda/2$ is split from the g.s. Kramers doublet by $3\lambda/2$. This level is four-fold degenerate with the atomic spin-orbit structure
\begin{equation}\label{eq:psi14}
	\begin{split}
\psi_1=&(\sqrt{2}d_+\uparrow+2d_3\downarrow)/\sqrt{6},\\
\psi_2=&(\sqrt{2}d_-\downarrow-2d_3\uparrow)/\sqrt{6},\\
\psi_3=&d_{+}\downarrow,\ \ \psi_4=d_{-}\uparrow.
\end{split}
\end{equation}
We note that $\psi_\pm$ are ground states of $\hat{H}_{SO}$ on the full space of all five d-orbitals since they minimize the SOC energy. In contrast,  the $j=1/2$ quadruplet are eigenstates of only the projection of $\hat{H}_{SO}$ on the $t_{2g}$ subspace, since only the $j=5/2$ and $j=3/2$ states are eigenstates of this Hamiltonian on the full space of five d-orbitals.
  
We use the perturbation theory to analyze the effects of hopping~\cite{landau1977quantum}. At $k\ll a^{-1}\sqrt{\lambda/t}$, hopping-induced mixing between the lowest-energy doublet $\psi_\pm$, and the $j=1/2$ quadruplet is negligible. The hopping term is diagonal on the subspace of plane waves 
\begin{equation}\label{eq:psi_sigma_Bloch}
\psi^{(0)}_{\vec{k},\sigma}=\frac{1}{\sqrt{N}}\sum_{\vec{n}} e^{ia\vec{k}\vec{n}}\hat{c}^+_{\vec{n},\sigma}|0\rangle=\hat{c}^+_{\vec{k},\sigma}|0\rangle,
\end{equation}
where $\sigma=\pm$ is pseudo-spin, the operator $\hat{c}^+_{\vec{n},\sigma}$ creates an electron in the state $\psi_\sigma$ on site $\vec{n}$, and operator $\hat{c}^+_{\vec{k},\sigma}$ creates an electron in the corresponding Bloch state.

The dispersion $\epsilon_\sigma(k)=[\frac{2}{3}t(ka)^2-\lambda]$ of the Bloch states Eq.~(\ref{eq:psi_sigma_Bloch}) is isotropic and independent of $\sigma$, with effective mass $m^*=3\hbar^2/4ta^2$. In contrast, the sub-bands derived from the $j=1/2$ states are anisotropic. For instance, the bands derived from $\psi_{1,2}$ are characterized by the effective masses $m_1=3\hbar^2/5ta^2$ along the $k_x$ and $k_y$ axes, and $m_3=3\hbar^2/2ta^2$ along the $k_z$ axis.

The degeneracy of the sub-bands derived from $\psi_{\sigma}$ is protected by a combination of time reversal and spatial inversion symmetries~\cite{dresselhaus2007group}. The latter is preserved by the antiferrodistortive transition but is lifted by the static ferroelectric distortions at heavy doping, or dynamic distortions associated with quantum paraelectricity at cryogenic temperatures. The effects of this symmetry breaking are discussed in the next section.

\begin{figure}
	\centering
	\includegraphics[width=\columnwidth]{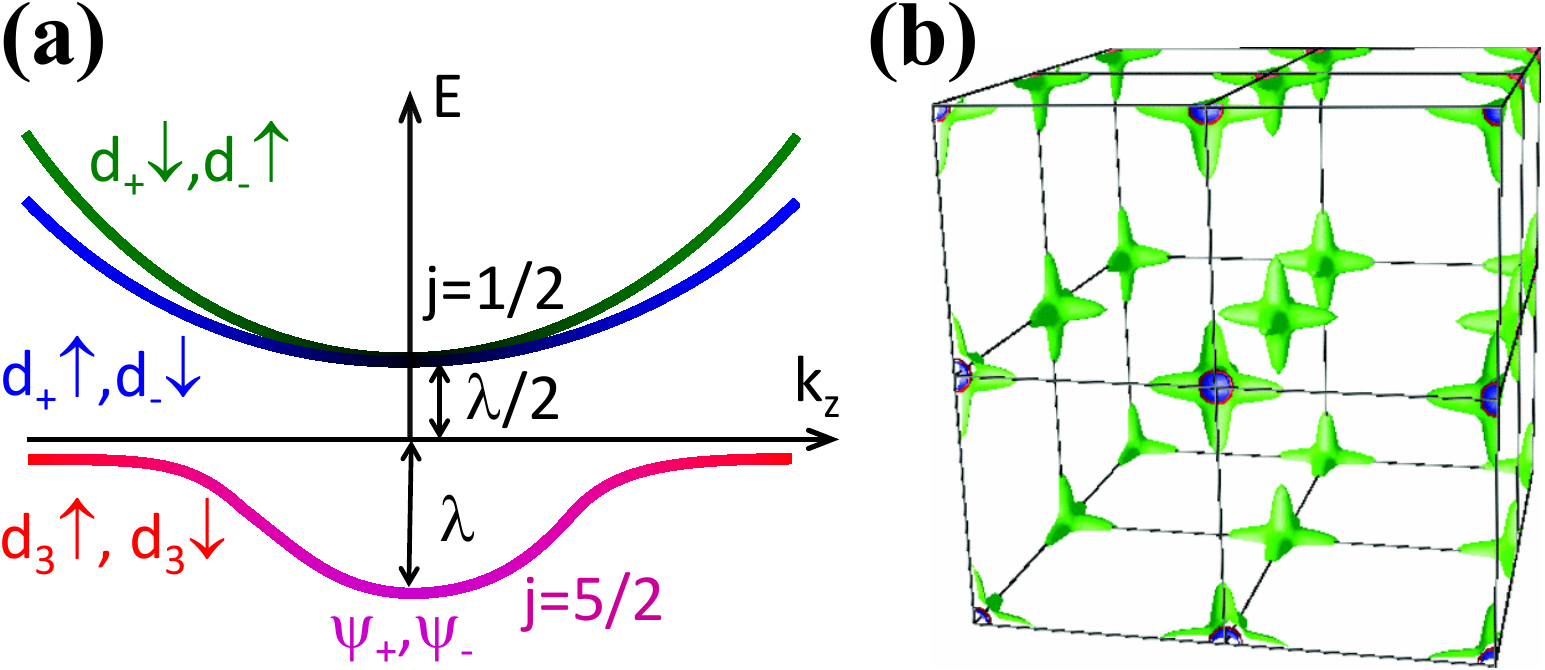}
	\caption{\label{fig:bands_SOC} (a) Structure of SOC conduction band along the $k_z$ axis. (b) Calculated SOC Fermi surface at $2\%$ doping ($n=3\times10^{20}$~cm$^{-3}$) reproduced from Ref.~\cite{PhysRevB.84.205111}.}
\end{figure}

Our analysis of SOC band structure is consistent with the observation of a quasi-isotropic Fermi surface at light doping~\cite{PhysRevX.3.021002}. The measured $m^*=1.6\times10^{-30}$~kg allows us to estimate $t=0.2$~eV. Hopping-induced orbital hybridization, which results in orbital moment quenching, is expected to onset around $k_F=a^{-1}\sqrt{\lambda/t}$, where $k_F=\sqrt[3]{3\pi^2n}$ is the Fermi wave vector, corresponding to doping $n\approx10^{19}$~cm$^{-3}$. At heavier doping, increased kinetic contribution results in the emergence of anisotropic band structure, as obtained in Section~\ref{sec:1p} in the limit of negligible SOC. The first-order hopping correction to $\psi^{(0)}_{\vec{k},\sigma}$ is
\begin{equation}\label{eq:chiral1}
	\psi^{(1)}_{\vec{k},\sigma}=-\sum_{n}\frac{\langle\psi^{(0)}_{\vec{k},n}|\hat{H}_{hop}|\psi^{(0)}_{\vec{k},\sigma}\rangle}{\epsilon_n-\epsilon_\sigma}\psi^{(0)}_{\vec{k},n},
\end{equation}
where $\psi^{(0)}_{\vec{k},n}$ are Bloch states derived from the orbitals $\psi_n$. For $\sigma=+$, the only finite matrix element of $\hat{H}_{hop}$ in this expression corresponds to $n=1$, and for $\sigma=-$ it is $n=2$. To the lowest order in $k$,
\begin{equation}\label{eq:f}
	\begin{split}
		&\langle\psi^{(0)}_{\vec{k},1}|\hat{H}_{hop}|\psi^{(0)}_{\vec{k},+}\rangle=\langle\psi^{(0)}_{\vec{k},2}|\hat{H}_{hop}|\psi^{(0)}_{\vec{k},-}\rangle\\
		&=\frac{ta^2}{3\sqrt{2}}(2k^2_z-k^2_x-k^2_y),
	\end{split}
\end{equation}
and $\epsilon_n-\epsilon_\sigma\approx3\lambda/2$. Based on these expressions, we infer that for $k\gtrsim a^{-1}\sqrt{\lambda/t}$ along $k_z$, $\psi_\sigma$ are mixed by hopping with other spin-orbital states, evolving at large $k$ into $d_3\uparrow$, $d_3\downarrow$ [Fig.~\ref{fig:bands_SOC}(a)]. Meanwhile, the $j=1/2$ quadruplet evolves into two $d_\pm$ sub-bands discussed above in the large-$k$ limit. By symmetry, the dependences are similar for other principal directions.

{\it Ab initio} calculations for $n=3\times10^{20}$~cm$^{-3}$ predicted a highly anisotropic star-shaped Fermi surface stretched along the principal axes [Fig.~\ref{fig:bands_SOC}(b)], consistent with this analysis~\cite{PhysRevB.84.205111}. The same calculations also showed that at heavier doping, the Fermi surface develops narrow lines along the principal axes spanning the Brillouin zone, also in agreement with our results.

\section{Effects of electron interactions}\label{sec:interactions}

In this section, we analyze the effects of the electronic properties discussed in the previous sections on electron correlations. The latter underlie one of the most puzzling properties of STO - sc observed at very low electron densities $n\gtrsim3\times10^{17}-10^{21}$~cm$^{-3}$, with the maximum transition temperature $T_c\approx0.4$~K at optimal doping $n\approx 10^{19}-10^{20}$~cm$^{-3}$~\cite{PhysRev.163.380,doi:10.1146/annurev-conmatphys-031218-013144}. Intriguingly, the optimal-doping regime corresponds to the emergence of a highly-anisotropic star-shaped Fermi surface, which involves an interplay of SOC and anisotropic orbitally-selective hopping beyond the usual isotropic single-band approximation of the BCS theory, highlighting the complexity of sc mechanisms in STO. 

Additional complexity is associated with the incipient ferroelectricity and other static or dynamic cubic symmetry breaking, such as strain in thin films, which was shown to strongly influence the superconducting properties~\cite{Ahadi2019,PhysRevLett.115.247002,PhysRevResearch.4.013019} and may be central to the mechanism of sc in STO~\cite{PhysRevB.97.144506,Rischau2017, Rowley2014,kedem2020paradigm}. Cubic symmetry breaking can lift the degeneracy of three $t_{2g}$ orbitals, resulting in a highly anisotropic electronic structure dominated by only one of these orbitals~\cite{dresselhaus2007group}. SOC can then induce spin anisotropy of these states, as observed in STO heterostructures~\cite{omar2021large}, leading to a complex spin-anisotropic and orbital-dependent electronic structure.

Here, we focus on the possible connection with unconventional sc~\cite{PhysRevB.101.100503,PhysRevLett.115.247002}. The latter term generally refers to sc not described by the BCS theory but is most often used for correlated metals - materials such as doped Mott insulators exemplified by cuprate HTSCs, whose anomalous electronic properties are determined by interactions~\cite{RevModPhys.60.585}. The Mott insulator state of undoped parent compounds of HTSCs is stabilized by electron interactions in a half-filled band, usually resulting in antiferromagnetic (AFM) ordering~\cite{Mott1949-pz}. Ordering is suppressed by doping, resulting in the metallic state characterized by residual AFM-coupled singlet pair correlations (Mott singlets). In the resonating valence bond (RVB) model of HTSCs, the latter mediate sc similarly to the Cooper pairs~\cite{doi:10.1126/science.235.4793.1196}.

Is it possible that STO exhibits similar interaction-induced singlet correlations, even though its conduction band is very far from half-filling at doping levels relevant to sc? We address this question using the Hubbard interaction Hamiltonian~\cite{pavarini2016quantum,de_medici_2017}

\begin{widetext}
	\begin{equation}\label{eq:H_int}
		\begin{split}
\hat{H}_{int}=&U\sum_{\vec{n},m} \hat{n}_{\vec{n},m,\uparrow}\hat{n}_{\vec{n},m,\downarrow}+(U-2J)\sum_{\vec{n},m'\ne m}\hat{n}_{\vec{n},m,\uparrow}\hat{n}_{\vec{n},m',\downarrow}+(U-3J)\sum_{\vec{n},m'<m,s}\hat{n}_{\vec{n},m,s}\hat{n}_{\vec{n},m',s}\\
+&J\sum_{\vec{n},m'\ne m}(\hat{c}^{\dagger}_{\vec{n},m ,\uparrow}\hat{c}^{\dagger}_{\vec{n},m',\downarrow}\hat{c}_{\vec{n},m,\downarrow}\hat{c}_{\vec{n},m',\uparrow}
+\hat{c}^{\dagger}_{\vec{n},m ,\uparrow}\hat{c}^{\dagger}_{\vec{n},m,\downarrow}\hat{c}_{\vec{n},m',\downarrow}\hat{c}_{\vec{n},m',\uparrow}),
		\end{split}
	\end{equation}
\end{widetext}
The first term is the Mott's energy of electrons with opposite spins on the same orbital, the next two terms represent the spin-dependent Hund's energy of electrons on different orbitals, and the last two are the symmetry-imposed pair spin-flip and orbital-hopping terms, with common notations for the coefficients~\cite{pavarini2016quantum, PhysRevB.83.205112}. We estimate $U=9-10$~eV, $J=1.3$~eV based on the prior Hubbard modeling of Ti compounds~\cite{PhysRevB.79.235126}. 

We first consider the cubic symmetry approximation, and discuss the effects of its breaking below. At  $n<10^{19}$~cm$^{-3}$, the conduction band states are derived from the spin-orbit coupled Kramers doublet $\psi_{\sigma}$ with orbital composition almost independent of the wavevector $k$. The interaction Hamiltonian Eq.~(\ref{eq:H_int}) projected on the states $\psi_\sigma$ with atomic structure Eq.~(\ref{eq:psi_sigma}) is
\begin{equation}\label{eq:H_int2}
	\hat{H}_{int}=V\sum_{\vec{n}}\hat{c}^+_{\vec{n},+}\hat{c}^+_{\vec{n},-}\hat{c}_{\vec{n},-}\hat{c}_{\vec{n},+},
\end{equation}
where $V=U-2J\approx 7$~eV and $\hat{c}_{\vec{n},\sigma}$ annihilates electron in the state $\psi_\sigma$ on site $\vec{n}$. Equation~(\ref{eq:H_int2}) is similar to the Mott's energy, with spin replaced by the pseudo-spin $\sigma$ and the Mott energy renormalized by the spin-orbit-coupled composition of $\psi_\sigma$.

To analyze the effects of interaction on electron correlations, we consider a volume $\Omega=L^3$, where $L=\sqrt[3]{N}a$ is the linear dimension, selected so that it contains just two electrons in the conduction band. The kinetic energy is minimized in the Bloch state with momentum $k=0$ for both electrons. According to the Pauli principle, they must occupy both states $\psi_+$ and $\psi_-$, so the interaction energy is $E_{int}=V/N=v$. On the other hand, if the two electrons are localized on different sites, their interaction energy vanishes, but their kinetic energy increases by $4t$. Thus, interaction can lead to instability of the Bloch states, as in the Mott mechanism~\cite{RevModPhys.40.677,PhysRevB.17.2575}.

In the reciprocal space, the onset of instability is expected to be manifested by finite  wavefunction components with the smallest possible nonzero wavevector, $\vec{k}_{\vec{l}}=2\pi\vec{l}/L$ for one of the two electrons, allowing the electrons to reduce their spatial overlap while minimizing the increase of kinetic energy. Here, $\vec{l}$ is a unit vector along one of the principal axes.

Consider a trial two-electron wavefunction~\cite{note}
\begin{equation}\label{eq:Psi_corr}
	\Psi=(\alpha\hat{c}^+_{0,+}\hat{c}^+_{0,-}+\beta_+\hat{c}^+_{0,+}\hat{c}^+_{\vec{k}_{\vec{l}},-}+\beta_-\hat{c}^+_{0,-}\hat{c}^+_{\vec{k}_{\vec{l}},+})|0\rangle,
\end{equation}
where $\hat{c}^+_{\vec{k},\sigma}$ creates a Bloch wave derived from $\psi_{\sigma}$, and $|\alpha|^2+|\beta_+|^2+|\beta_-|^2=1$. The kinetic energy is 
\begin{equation}\label{eq:E_kin}
	E_{kin}=\epsilon_1(|\beta_+|^2+|\beta_-|^2),
\end{equation}
where $\epsilon_1=2\hbar^2\pi^2/m^*L^2$ is the energy of the $\vec{k}_{\vec{l}}$ state relative to the $k=0$ state.

The interaction energy obtained by transforming Eq.~(\ref{eq:Psi_corr}) into the coordinate representation is 
\begin{equation}\label{eq:E_int}
	E_{int}=v[1-2Re(\beta_+\beta_-^*)].
\end{equation}
Finite $\beta_+$, $\beta_-$ increase $E_{kin}$ and decrease $E_{int}$. The latter is minimized in the pseudo-spin singlet state with $\beta_+=\beta_-=\beta e^{i\phi}$, where $\beta$ is real and $\phi$ describes the gauge symmetry of the singlet. The total energy is
\begin{equation}\label{eq:E_tot}
E=E_{kin}+E_{int}=v+2\beta^2(\epsilon_1-v).
\end{equation}
At $v<\epsilon_1$, this energy is minimized in the Bloch state with $\alpha=1$, $\beta=0$, while at $v>\epsilon_1$ it is minimized in the pseudo-spin singlet state with $\alpha=0$, $\beta=1/\sqrt{2}$. This is a first-order transition since the order parameter defined by the singlet amplitude varies discontinuously across the transition.

Including other wavevectors $\vec{k}_{\vec{l}}$ in the trial wavefunction does not affect the energy of the singlet state, resulting in the degeneracy of the singlet with respect to the distribution among these wavevectors,
\begin{equation}\label{eq:singlet}
	\Psi_s=\sum_{\vec{l}}\beta_{\vec{l}}(\hat{c}^+_{0,+}\hat{c}^+_{\vec{k}_{\vec{l}},-}+\hat{c}^+_{0,-}\hat{c}^+_{\vec{k}_{\vec{l}},+})|0\rangle,
\end{equation}
limited only by the normalizing condition $\sum|\beta_{\vec{l}}|^2=1/2$. Due to the finite wavevector, a singlet that involves a particular value of $\vec{k}_{\vec{l}}$ has a finite velocity
\begin{equation}\label{eq:jl}
	\vec{v}_{\vec{l}}=\frac{1}{\hbar}\frac{d\epsilon(k)}{d\vec{k}}|_{\vec{k}=\vec{k}_{\vec{l}}}=\frac{2^{2/3}\hbar\pi n^{4/3}}{m^*}\vec{l},
\end{equation}
which carries current density $\vec{j}=e\vec{v}_{\vec{l}}/Na^3$. Thus, this g.s. can carry current due to its momentum-degeneracy, i.e., it is not an insulator but a metal in which charge current is carried by pseudo-spin singlets similar to the Mott singlets in the metallic state of doped Mott insulators~\cite{doi:10.1126/science.235.4793.1196}. This state can be a correlated metal like HTSCs above their $T_c$ or a superconductor if the pairs exhibit long-range coherence. We leave the analysis of these possibilities to future studies.

In our model, the volume $\Omega=L^3$ is determined by the condition that it contains two electrons, or equivalently $L=\sqrt[3]{2/n}$, where $n$ is the electron density. Using the effective tight-binding bandwidth $w=2\hbar^2/m^*a^2$, the instability criterion $v>\epsilon_1$ can be written in a simple form $n>n_c=2(\pi^2w/aV)^3$, or equivalently, the critical doping level is $d_c=2(\pi^2w/V)^3$.

Using the experimental value $m^*=1.6\times10^{-30}$~kg~\cite{PhysRevX.3.021002}, we obtain  $w=0.5$~eV, resulting in an unphysically large  critical doping $d_c\approx 50\%$. Nevertheless, we now show that the proposed correlation mechanism is relevant to STO due to the orbitally-selective hopping anisotropy.

\begin{figure}
	\centering
	\includegraphics[width=0.9\columnwidth]{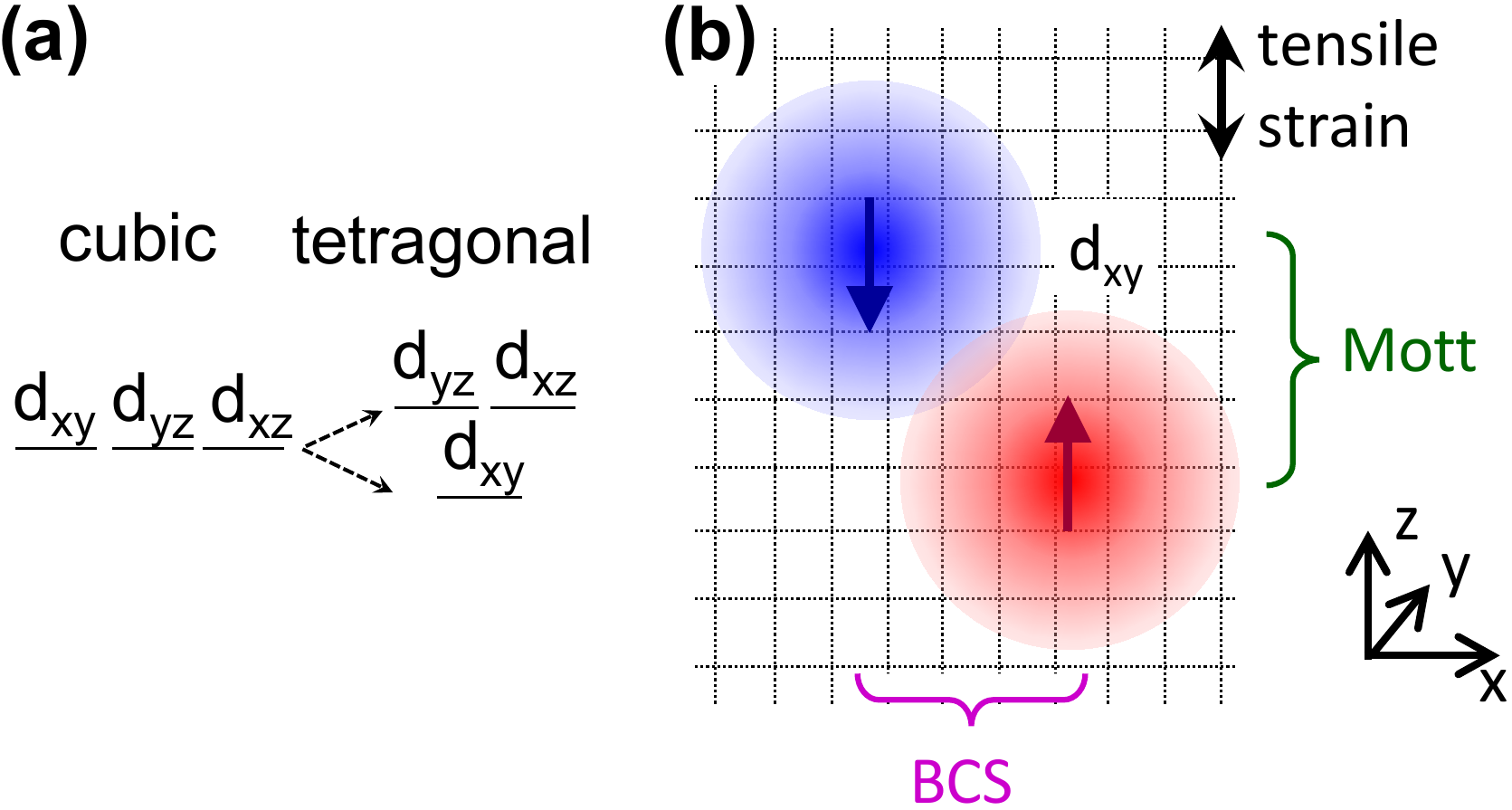}
	\caption{\label{fig:Mott_BCS} (a) Splitting of the $t_{2g}$ levels by tetragonal distortion. (b) Possible coexistence of Cooper pairing and Mott correlations in a cubic symmetry-broken state dominated by one of the $t_{2g}$ orbitals.}
\end{figure}

\textbf{Effects of cubic symmetry breaking.}  Ample evidence suggests that incipient ferroelectric distortions and lattice strain are important for sc in STO~\cite{Ahadi2019,PhysRevLett.115.247002,PhysRevResearch.4.013019}. We now argue that they are also important for the Mott-like singlet correlations discussed above. Consider, for example, tensile strain in the z-direction, which reduces the hopping amplitude in this direction, resulting in the splitting of the $t_{2g}$ manifold into the lower-energy $d_{xy}$ sub-band and higher-energy $d_{xz}$, $d_{yz}$ sub-bands [Fig.~\ref{fig:Mott_BCS}(a)]. It may be induced by epitaxy in thin films, or associated with incipient ferroelectricity in the bulk, which at low temperature can be treated adiabatically due to the critical slowdown of ferroelectric fluctuations. A similar splitting is expected due to the symmetry breaking at the interfaces, as was shown experimentally and confirmed by band structure calculations~\cite{King2014,Walker2015}.

Hopping in the $d_{xy}$ sub-band is mostly confined to the xy-plane, so this sub-band is almost non-dispersive in the z direction, i.e., the corresponding component $m^*_3$ of the effective mass tensor almost diverges. In this approximation, the kinetic energy cost $\epsilon_1$ associated with the singlet correlation, which is inversely proportional to the effective mass, becomes strongly dependent on the direction of wavevector $\vec{k}_{\vec{l}}$, vanishing for $\vec{k}_{\vec{l}}$ collinear with the z-axis. The only condition for the correlation is then that such wavevectors exist, which is satisfied for $L>2a$, i.e., for the doping levels $d<25\%$. Qualitatively, this result can be viewed as a special anisotropic case of enhanced correlations in  flat bands extensively discussed in the context of twisted multilayer graphene~\cite{Balents2020-gp}. 

A small but finite dispersion in the z-direction is expected due to SOC or polar symmetry breaking that reduces orbital selectivity of hopping, which can be approximated by a large $m^*_3$. Using the same analysis as above with the effective Mott parameter $V$ replaced by the Mott energy $U$ for the single orbital, we obtain the criterion for the proposed Mott-like correlation
\begin{equation}\label{eq:Mott}
n>\frac{16\pi^6\hbar^6}{m^{*3}_3U^3a^9},
\end{equation}
which shows that the critical carrier density for the onset of the correlation exhibits a strong dependence on the dominant component of the effective mass tensor. We estimate the critical value $m^*_3=20m^*$ required for the onset of the proposed correlations at the carrier density $n=3\times10^{17}$~cm$^{-3}$. This value is plausible judging from the large sub-band anisotropy obtained in realistic band structure calculations~\cite{PhysRevB.84.205111,PhysRevB.6.4718}. 

The proposed correlation mechanism is also likely relevant in the limit of negligible effects of cubic symmetry breaking, in the doping regime $n>10^{19}$~cm$^{-3}$ where a large  conduction band anisotropy emerges, as manifested by the start-shaped Fermi surface in Fig.~\ref{fig:bands_SOC}(b). The above analysis can be extended to show that Mott correlations can be stabilized by interactions between electrons on the Fermi surface with momenta along  the principal direction characterized by a large effective mass. We leave a detailed analysis of this limit to future studies.

\section{Mott vs BCS correlations}\label{sec:sc}

In this section, we discuss the possible interplay between Mott and BCS correlations. We start with the isotropic model of pseudo-spin Mott correlations discussed in the previous section. 
Consider two electrons characterized by the wavevectors $\vec{k}_1$, $\vec{k}_2$ and opposite pseudospins $\sigma_2=-\sigma_1$. The wavefunction of their uncorrelated two-electron Bloch state is $\Psi_2=\hat{c}^+_{\vec{k}_1,\sigma_1}\hat{c}^+_{\vec{k}_1,\sigma_2}|0\rangle$, and the corresponding interaction energy is $v$. The latter vanishes in the singlet state 
\begin{equation}\label{eq:singlet2}
	\Psi_s=\frac{1}{\sqrt{2}}(\hat{c}^+_{\vec{k}_1,+}\hat{c}^+_{\vec{k}_2,-}+\hat{c}^+_{\vec{k}_1,-}\hat{c}^+_{\vec{k}_2,+})|0\rangle.
\end{equation}
This expression is analogous to Eq.~(\ref{eq:singlet}), demonstrating that repulsion-induced singlet correlations reduce interaction energy for any pair of wavevectors. This result can be contrasted with Cooper pairing, which is limited to opposite momenta~\cite{tinkham2004introduction}. Qualitatively, this difference can be interpreted as a consequence of the fact that two electrons can avoid each other in many different ways, while their proximity in a bound Cooper pair limits their possible motion.

The two-electron wavefunction describing a Cooper pair is~\cite{PhysRev.104.1189}
\begin{equation}\label{eq:singlet_BCS}
	\Psi_s=\frac{1}{\sqrt{2}}(\hat{c}^+_{\vec{k},\uparrow}\hat{c}^+_{-\vec{k},\downarrow}-\hat{c}^+_{\vec{k},\downarrow}\hat{c}^+_{-\vec{k},\uparrow})|0\rangle,
\end{equation}
where $|0\rangle$ denotes the g.s. of the Fermi gas. This expression has the same form as Eq.~(\ref{eq:singlet2}) with $\vec{k}_1=\vec{k}$ and $\vec{k}_2=-\vec{k}$, except for the opposite relative signs of the two wavefunction components. This difference is consistent with the difference between the two mechanisms of correlations. Mott correlations result from repulsive interaction. Accordingly, the positive sign in Eq.~(\ref{eq:singlet2}) minimizes electron overlap. In contrast, BCS correlations are driven by attractive interaction, with the negative sign in Eq.~(\ref{eq:singlet_BCS}) maximizing electron overlap.

This difference seems to suggest that BCS correlations are incompatible with Mott correlations. However, the differences between the wavevector- and directional-dependences of the two correlations allow them to coexist, as illustrated in Fig.~\ref{fig:Mott_BCS}(b) for the broken-symmetry state discussed in the previous section. In the presence of tensile strain in the z-direction, the $d_{xy}$ subband dominates the lowest-energy conduction band states. Hopping in the z-direction is suppressed, which can result in Mott correlation for the components of the two-electron wavefunction with wavevectors in the z-direction corresponding to large effective mass, while BCS correlations may be dominant for other wavevectors corresponding to small effective mass. Anisotropic correlations may be particularly important for sc at small carrier concentrations since the group velocity for the momenta in the direction of large effective mass is small, making the usual retarded phonon-mediated effective electron attraction mechanism of the Migdal-Eliashberg theory inapplicable. 

\section{Summary}\label{sec:discussion}

In this work, we used tight-binding analysis of the conduction band in STO to show that in the cubic-symmetry approximation, the lowest-energy electronic states are dominated by spin-orbit coupling, which results in the formation of a spin-orbit Kramers doublet that mixes all three $t_{2g}$ orbitals of Ti and is characterized by a large moment $j=5/2$. Meanwhile, higher-energy conduction states that become populated at carrier densities $n>10^{19}$~cm$^{-3}$ involve three sub-bands characterized by highly anisotropic orbitally-selective hopping, with very large effective masses along the principal axes. This result is supported by the observation of effective mass increase with increasing temperature consistent with the increase of thermal population of anisotropic higher-energy states identified in our work~\cite{PhysRevX.10.031025}. Cubic symmetry breaking associated with static lattice distortions, interfaces, or the dynamic effects of incipient ferroelectricity can dramatically restructure the low-energy conduction states, resulting in highly anisotropic properties even at low carrier concentrations.

As a consequence of orbitally-selective hopping anisotropy, each of the three $t_{2g}$ sub-bands is nearly dispersionless along one of the principal axes, resulting in  interaction-induced electronic correlations analogous to those in flat bands~\cite{Balents2020-gp}. We utilized the Hubbard model to show that, at sufficiently high doping determined by the effective mass in this direction, the Bloch states become unstable with respect to Mott-like singlet correlations. Finally, we argued that Mott correlations may coexist with the BCS correlations due to their different wavevector dependence. 

Anisotropic orbitally-selective electron hopping may be relevant to the properties of high-mobility 2d electron gas at STO/LaAlO$_3$ interfaces~\cite{Ohtomo2004}, where cubic symmetry breaking results in the dominance of the $d_{xy}$ orbital~\cite{King2014,Walker2015}. According to our analysis, electron hopping is then confined to the plane of the interface, avoiding electron scattering from the interface imperfections. We note that SOC is expected to result in a perpendicular spin anisotropy, which can be tested by electron spin resonance (ESR) or Hanle measurements of the dynamics of spins injected into the 2d gas. The large orbital moments of the conduction states predicted by our analysis for the cubic phase may be manifested by an anomalously low Land\'e factor in the ESR of the conduction band of bulk STO. To the best of our knowledge, only impurity ESR has been studied so far~\cite{Glinchuk2001}. 

The possibility that Mott singlet correlations may contribute to superconductivity in STO may provide a simple interpretation for the dome-like dependence of critical temperature on doping. At low doping, the criterion Eq.~(\ref{eq:Mott}) for Mott correlations is satisfied only for a small region of the Fermi surface in one of the principal directions, which expands with increased doping, resulting in increasing $T_c$. At heavy doping, the Mott correlations mechanism becomes suppressed by multi-band contributions. The dependence of the identified Mott correlations on the band structure anisotropy also provides a natural interpretation for the strong correlation between superconducting properties and lattice distortions.

A possible contribution of Mott correlations to superconductivity in STO connects this material to high-temperature superconductors, as envisioned by K.A. M\"{u}ller~\cite{condmat5040060}. Our analysis indicates that multi-band effects and spin-orbit interaction play a significant role in STO, suggesting a connection to other multi-band unconventional superconductors characterized by significant spin-orbit effects, including pnictides, twisted multilayer graphene, and superconducting ruthenates.

Band structure calculations by A.M. and S.U. were supported by the DOE BES Award \# DE-SC0018976. Analysis of correlations by S.U. and E.T. was supported by NSF Awards ECCS-1804198 and ECCS-2005786.

\section*{Appendix: Alternative derivation of interaction-induced singlet correlation}

Here, we analyze the conditions for instability of the Bloch states of two electrons induced by repulsive interactions and determine the resulting correlated state without relying on the {\it ad hoc} trial wavefunction Eq.~(\ref{eq:Psi_corr}). We assume cubic symmetry, but with minor modifications the analysis is also applicable to anisotropic systems, such as STO in the presence of ferroelectric distortions or strain. We start by transforming the interaction Hamiltonian Eq.~(\ref{eq:H_int2}) into the reciprocal space 
\begin{equation}\label{eq:H_int_k}
	\hat{H}_{int}=v\sum_{\vec{k},\vec{k}',\vec{q}}\hat{c}^+_{\vec{k}+\vec{q},+}\hat{c}^+_{\vec{k}'-\vec{q},-}\hat{c}_{\vec{k}',-}\hat{c}_{\vec{k},+}.
\end{equation}

We consider the contributions of zero wavevectors and the smallest finite wavevectors, $k_{\vec{l}}$. The first contribution to the interaction energy is provided by the term with $k=k'=q=0$. The remaining terms include those with two zero and two finite wavevectors, and those with here all four finite wavevectors. The latter are characterized by large kinetic energy and can be neglected sufficiently close to the instability. The remaining terms can be classified into three groups based on which pair of wavevectors have zero values. The first group with the form $\hat{c}^+_{0,+}\hat{c}^+_{0,-}\hat{c}_{\vec{q},-}\hat{c}_{-\vec{q},+}$ can be neglected, since it requires that both electrons are in the finite-momentum state. The remaining interaction terms are
\begin{equation}\label{eq:H_int_k2}
	\begin{split}
		\hat{H}_{int}=&v\hat{n}_{0,+}\hat{n}_{0,-}\\
		+&v\sum_{\vec{l},\sigma}(\hat{n}_{0,\sigma}\hat{n}_{\vec{k}_{\vec{j}},-\sigma}+\hat{c}^+_{0,\sigma}\hat{c}^+_{\vec{k}_{\vec{l}},-\sigma}\hat{c}_{0,-\sigma}\hat{c}_{\vec{k}_{\vec{l}},\sigma}).
	\end{split}
\end{equation}
The first two terms describe the interaction energy associated with the uncorrelated populations of zero- and finite-momentum states, while the last term describes the effect of correlations between these states that allow electrons to reduce the interaction energy. This equation can be written in an alternative form
\begin{equation}\label{eq:H_int_k3}
		\hat{H}_{int}=v(\hat{n}_{0,+}\hat{n}_{0,-}+\sum_{\vec{l},\sigma}\hat{n}_{0,\sigma}\hat{n}_{\vec{k}_{\vec{l}},-\sigma}-\sum_{\vec{l}}\hat{b}^+_{\vec{l}}\hat{b}_{\vec{l}}),
\end{equation}
where the operator
\begin{equation}\label{eq:b}
	\hat{b}_{\vec{l}}=\hat{c}_{0,+}\hat{c}_{\vec{k}_{\vec{l}},-}+\hat{c}_{0,-}\hat{c}_{\vec{k}_{\vec{l}},+},
\end{equation}
annihilates a singlet pair described by Eq.~(\ref{eq:singlet}). 

The condition that there are two electrons in volume $\Omega$ can be written as 
\begin{equation}\label{eq:population}
	\hat{n}_{0,+}+\hat{n}_{0,-}+\sum_{\vec{l},\sigma}\hat{n}_{\vec{k}_{\vec{l}},\sigma}=2,
\end{equation}
which allows us to rewrite the first term in Eq.~(\ref{eq:H_int_k3}) as 
\begin{equation}\label{eq:nono}
	\hat{n}_{0,+}\hat{n}_{0,-}=\frac{1}{2}(\hat{n}_{0,+}+\hat{n}_{0,-}-\sum_{\vec{l},\sigma}\hat{n}_{\vec{k}_{\vec{l}},\sigma}),
\end{equation}
where we have used the identity $\hat{n}_{\vec{k},\sigma}^2=\hat{n}_{\vec{k},\sigma}$ and neglected the terms which require that both electrons are in finite-wavevector states. 

In the absence of magnetism, the electrons are equally distributed between two pseudo-spins, 
\begin{equation}\label{eq:population}
	\hat{n}_{0,\sigma}+\sum_{\vec{l}}\hat{n}_{\vec{k}_{\vec{l}},\sigma}=1.
\end{equation}
This relation allows us to transform the second term in Eq.~(\ref{eq:H_int_k3}) into
\begin{equation}\label{eq:nono}
	\sum_{\vec{l},\sigma}\hat{n}_{0,\sigma}\hat{n}_{\vec{k}_{\vec{l}},-\sigma}=\sum_{\vec{l},\sigma}\hat{n}_{\vec{k}_{\vec{l}},-\sigma},
\end{equation}
where we have again neglected the term requiring that both electrons are in finite-wavevector states.

The resulting interaction Hamiltonian is
\begin{equation}\label{eq:H_int_k4}
	\hat{H}_{int}=\frac{v}{2}\sum_{\vec{l},\sigma}(\hat{n}_{0,\sigma}+\hat{n}_{\vec{k}_{\vec{l}},-\sigma})-v\sum_{\vec{l}}\hat{b}^+_{\vec{l}}\hat{b}_{\vec{l}}.
\end{equation}

In the mean-field approach, singlet correlations can be described by the gauge-symmetric multi-component order parameter 
\begin{equation}\label{eq:correlator1}
	\begin{split}
		\Delta_{\vec{l}}=v\langle \hat{b}_{\vec{l}}\rangle=v\langle \hat{c}_{0,+}\hat{c}_{\vec{k}_{\vec{l}},-}+\hat{c}_{0,-}\hat{c}_{\vec{k}_{\vec{l}},+}\rangle.
	\end{split}
\end{equation}
This order parameter can be utilized to reduce the interaction Hamiltonian Eq.~(\ref{eq:H_int_k4}) to a quadratic form with respect to fermionic operators, as in the mean-field RVB approximation~\cite{BASKARAN1987973}. Detailed analysis will be presented elsewhere.

Here, we analyze the properties of the Hamiltonian without linearization. The kinetic energy contribution to the Hamiltonian projected on the subspace of states with wavevectors $k=0$ and $\vec{k}_{\vec{l}}$ is
\begin{equation}\label{eq:H_kin}
	\hat{H}_{kin}=\epsilon_1\sum_{\vec{l},\sigma}\hat{n}_{\vec{k}_{\vec{l}},\sigma}.
\end{equation}
The total Hamiltonian can be written as
\begin{equation}\label{eq:H_tot}
	\hat{H}=\hat{H}_{kin}+\hat{H}_{int}=\hat{H}_1+\hat{H}_2,
\end{equation}
where $\hat{H}_1$ and $\hat{H}_2$ are effective single-particle and two-particle Hamiltonians,
\begin{equation}
	\begin{split}
		\hat{H}_1=&\frac{v}{2}(\hat{n}_{0,+}+\hat{n}_{0,-})+\sum_{\vec{l},\sigma}(\epsilon_1+\frac{v}{2})\hat{n}_{\vec{k}_{\vec{l}},\sigma},\\
		\label{eq:H_12}
		\hat{H}_2=&-v\sum_{\vec{l}}\hat{b}^+_{\vec{l}}\hat{b}_{\vec{l}}.
	\end{split}
\end{equation}

$\hat{H}_1$ commutes with $\hat{H}_2$, and therefore they have a common basis of eigenstates, which are the stationary states of the full Hamiltonian. The two eigenstates of $\hat{H}_2$ on the space of states of two electrons are, i) the state with no singlets, and ii) the state with one singlet arbitrarily distributed over different $\vec{l}$. The degeneracy arises because singlet energy is independent of $\vec{l}$. 
The total energy of the state with one singlet is $E_s=\epsilon_1$. In the state with no singlets, the g.s. of $\hat{H}_1$ is the Bloch state of two electrons with $k=0$, and the total energy $E_0=v$. Thus, at $v>\epsilon_1$ the system experiences a transition from the Bloch g.s. to the singlet g.s. For anisotropic dispersion, the energy $\epsilon_1$ become dependent on $\vec{l}$. The instability is expected to occur when the condition $v>\epsilon_1(\vec{l})$ is satisfied for the smallest $\epsilon_1(\vec{l})$, i.e. for the axis corresponding to the largest effective mass. The resulting correlation involves electrons with wavevectors along this axis. These results are consistent with the analysis in Section~\ref{sec:interactions} based on the two-electron wavefunction ansatz Eq.~(\ref{eq:Psi_corr}).

\bibliography{SrTiO3}

\begin{thebibliography}{57}%
\makeatletter
\providecommand \@ifxundefined [1]{%
 \@ifx{#1\undefined}
}%
\providecommand \@ifnum [1]{%
 \ifnum #1\expandafter \@firstoftwo
 \else \expandafter \@secondoftwo
 \fi
}%
\providecommand \@ifx [1]{%
 \ifx #1\expandafter \@firstoftwo
 \else \expandafter \@secondoftwo
 \fi
}%
\providecommand \natexlab [1]{#1}%
\providecommand \enquote  [1]{``#1''}%
\providecommand \bibnamefont  [1]{#1}%
\providecommand \bibfnamefont [1]{#1}%
\providecommand \citenamefont [1]{#1}%
\providecommand \href@noop [0]{\@secondoftwo}%
\providecommand \href [0]{\begingroup \@sanitize@url \@href}%
\providecommand \@href[1]{\@@startlink{#1}\@@href}%
\providecommand \@@href[1]{\endgroup#1\@@endlink}%
\providecommand \@sanitize@url [0]{\catcode `\\12\catcode `\$12\catcode
  `\&12\catcode `\#12\catcode `\^12\catcode `\_12\catcode `\%12\relax}%
\providecommand \@@startlink[1]{}%
\providecommand \@@endlink[0]{}%
\providecommand \url  [0]{\begingroup\@sanitize@url \@url }%
\providecommand \@url [1]{\endgroup\@href {#1}{\urlprefix }}%
\providecommand \urlprefix  [0]{URL }%
\providecommand \Eprint [0]{\href }%
\providecommand \doibase [0]{https://doi.org/}%
\providecommand \selectlanguage [0]{\@gobble}%
\providecommand \bibinfo  [0]{\@secondoftwo}%
\providecommand \bibfield  [0]{\@secondoftwo}%
\providecommand \translation [1]{[#1]}%
\providecommand \BibitemOpen [0]{}%
\providecommand \bibitemStop [0]{}%
\providecommand \bibitemNoStop [0]{.\EOS\space}%
\providecommand \EOS [0]{\spacefactor3000\relax}%
\providecommand \BibitemShut  [1]{\csname bibitem#1\endcsname}%
\let\auto@bib@innerbib\@empty
\bibitem [{\citenamefont {Bednorz}\ and\ \citenamefont
  {M\"uller}(1988)}]{RevModPhys.60.585}%
  \BibitemOpen
  \bibfield  {author} {\bibinfo {author} {\bibfnamefont {J.~G.}\ \bibnamefont
  {Bednorz}}\ and\ \bibinfo {author} {\bibfnamefont {K.~A.}\ \bibnamefont
  {M\"uller}},\ }\href {https://doi.org/10.1103/RevModPhys.60.585} {\bibfield
  {journal} {\bibinfo  {journal} {Rev. Mod. Phys.}\ }\textbf {\bibinfo {volume}
  {60}},\ \bibinfo {pages} {585} (\bibinfo {year} {1988})}\BibitemShut
  {NoStop}%
\bibitem [{\citenamefont {Scheerer}\ \emph {et~al.}(2020)\citenamefont
  {Scheerer}, \citenamefont {Boselli}, \citenamefont {Pulmannova},
  \citenamefont {Rischau}, \citenamefont {Waelchli}, \citenamefont {Gariglio},
  \citenamefont {Giannini}, \citenamefont {van~der Marel},\ and\ \citenamefont
  {Triscone}}]{condmat5040060}%
  \BibitemOpen
  \bibfield  {author} {\bibinfo {author} {\bibfnamefont {G.}~\bibnamefont
  {Scheerer}}, \bibinfo {author} {\bibfnamefont {M.}~\bibnamefont {Boselli}},
  \bibinfo {author} {\bibfnamefont {D.}~\bibnamefont {Pulmannova}}, \bibinfo
  {author} {\bibfnamefont {C.~W.}\ \bibnamefont {Rischau}}, \bibinfo {author}
  {\bibfnamefont {A.}~\bibnamefont {Waelchli}}, \bibinfo {author}
  {\bibfnamefont {S.}~\bibnamefont {Gariglio}}, \bibinfo {author}
  {\bibfnamefont {E.}~\bibnamefont {Giannini}}, \bibinfo {author}
  {\bibfnamefont {D.}~\bibnamefont {van~der Marel}},\ and\ \bibinfo {author}
  {\bibfnamefont {J.-M.}\ \bibnamefont {Triscone}},\ }\href@noop {} {\bibfield
  {journal} {\bibinfo  {journal} {Condens. Matter}\ }\textbf {\bibinfo {volume}
  {5}},\ \bibinfo {pages} {60} (\bibinfo {year} {2020})}\BibitemShut {NoStop}%
\bibitem [{\citenamefont {Gastiasoro}\ \emph {et~al.}(2020)\citenamefont
  {Gastiasoro}, \citenamefont {Ruhman},\ and\ \citenamefont
  {Fernandes}}]{GASTIASORO2020168107}%
  \BibitemOpen
  \bibfield  {author} {\bibinfo {author} {\bibfnamefont {M.~N.}\ \bibnamefont
  {Gastiasoro}}, \bibinfo {author} {\bibfnamefont {J.}~\bibnamefont {Ruhman}},\
  and\ \bibinfo {author} {\bibfnamefont {R.~M.}\ \bibnamefont {Fernandes}},\
  }\href {https://doi.org/https://doi.org/10.1016/j.aop.2020.168107} {\bibfield
   {journal} {\bibinfo  {journal} {Annals of Physics}\ }\textbf {\bibinfo
  {volume} {417}},\ \bibinfo {pages} {168107} (\bibinfo {year} {2020})},\
  \bibinfo {note} {eliashberg theory at 60: Strong-coupling superconductivity
  and beyond}\BibitemShut {NoStop}%
\bibitem [{\citenamefont {Collignon}\ \emph {et~al.}(2019)\citenamefont
  {Collignon}, \citenamefont {Lin}, \citenamefont {Rischau}, \citenamefont
  {Fauqué},\ and\ \citenamefont
  {Behnia}}]{doi:10.1146/annurev-conmatphys-031218-013144}%
  \BibitemOpen
  \bibfield  {author} {\bibinfo {author} {\bibfnamefont {C.}~\bibnamefont
  {Collignon}}, \bibinfo {author} {\bibfnamefont {X.}~\bibnamefont {Lin}},
  \bibinfo {author} {\bibfnamefont {C.~W.}\ \bibnamefont {Rischau}}, \bibinfo
  {author} {\bibfnamefont {B.}~\bibnamefont {Fauqué}},\ and\ \bibinfo {author}
  {\bibfnamefont {K.}~\bibnamefont {Behnia}},\ }\href
  {https://doi.org/10.1146/annurev-conmatphys-031218-013144} {\bibfield
  {journal} {\bibinfo  {journal} {Annual Review of Condensed Matter Physics}\
  }\textbf {\bibinfo {volume} {10}},\ \bibinfo {pages} {25} (\bibinfo {year}
  {2019})},\ \Eprint
  {https://arxiv.org/abs/https://doi.org/10.1146/annurev-conmatphys-031218-013144}
  {https://doi.org/10.1146/annurev-conmatphys-031218-013144} \BibitemShut
  {NoStop}%
\bibitem [{\citenamefont {M\"uller}\ and\ \citenamefont
  {Burkard}(1979)}]{PhysRevB.19.3593}%
  \BibitemOpen
  \bibfield  {author} {\bibinfo {author} {\bibfnamefont {K.~A.}\ \bibnamefont
  {M\"uller}}\ and\ \bibinfo {author} {\bibfnamefont {H.}~\bibnamefont
  {Burkard}},\ }\href {https://doi.org/10.1103/PhysRevB.19.3593} {\bibfield
  {journal} {\bibinfo  {journal} {Phys. Rev. B}\ }\textbf {\bibinfo {volume}
  {19}},\ \bibinfo {pages} {3593} (\bibinfo {year} {1979})}\BibitemShut
  {NoStop}%
\bibitem [{\citenamefont {Uwe}\ and\ \citenamefont
  {Sakudo}(1976)}]{PhysRevB.13.271}%
  \BibitemOpen
  \bibfield  {author} {\bibinfo {author} {\bibfnamefont {H.}~\bibnamefont
  {Uwe}}\ and\ \bibinfo {author} {\bibfnamefont {T.}~\bibnamefont {Sakudo}},\
  }\href {https://doi.org/10.1103/PhysRevB.13.271} {\bibfield  {journal}
  {\bibinfo  {journal} {Phys. Rev. B}\ }\textbf {\bibinfo {volume} {13}},\
  \bibinfo {pages} {271} (\bibinfo {year} {1976})}\BibitemShut {NoStop}%
\bibitem [{\citenamefont {Burke}\ and\ \citenamefont
  {Pressley}(1971)}]{BURKE1971191}%
  \BibitemOpen
  \bibfield  {author} {\bibinfo {author} {\bibfnamefont {W.}~\bibnamefont
  {Burke}}\ and\ \bibinfo {author} {\bibfnamefont {R.}~\bibnamefont
  {Pressley}},\ }\href
  {https://doi.org/https://doi.org/10.1016/0038-1098(71)90115-3} {\bibfield
  {journal} {\bibinfo  {journal} {Solid State Communications}\ }\textbf
  {\bibinfo {volume} {9}},\ \bibinfo {pages} {191} (\bibinfo {year}
  {1971})}\BibitemShut {NoStop}%
\bibitem [{\citenamefont {Lin}\ \emph {et~al.}(2014{\natexlab{a}})\citenamefont
  {Lin}, \citenamefont {Bridoux}, \citenamefont {Gourgout}, \citenamefont
  {Seyfarth}, \citenamefont {Kr\"amer}, \citenamefont {Nardone}, \citenamefont
  {Fauqu\'e},\ and\ \citenamefont {Behnia}}]{PhysRevLett.112.207002}%
  \BibitemOpen
  \bibfield  {author} {\bibinfo {author} {\bibfnamefont {X.}~\bibnamefont
  {Lin}}, \bibinfo {author} {\bibfnamefont {G.}~\bibnamefont {Bridoux}},
  \bibinfo {author} {\bibfnamefont {A.}~\bibnamefont {Gourgout}}, \bibinfo
  {author} {\bibfnamefont {G.}~\bibnamefont {Seyfarth}}, \bibinfo {author}
  {\bibfnamefont {S.}~\bibnamefont {Kr\"amer}}, \bibinfo {author}
  {\bibfnamefont {M.}~\bibnamefont {Nardone}}, \bibinfo {author} {\bibfnamefont
  {B.}~\bibnamefont {Fauqu\'e}},\ and\ \bibinfo {author} {\bibfnamefont
  {K.}~\bibnamefont {Behnia}},\ }\href
  {https://doi.org/10.1103/PhysRevLett.112.207002} {\bibfield  {journal}
  {\bibinfo  {journal} {Phys. Rev. Lett.}\ }\textbf {\bibinfo {volume} {112}},\
  \bibinfo {pages} {207002} (\bibinfo {year} {2014}{\natexlab{a}})}\BibitemShut
  {NoStop}%
\bibitem [{\citenamefont {Lin}\ \emph {et~al.}(2015)\citenamefont {Lin},
  \citenamefont {Rischau}, \citenamefont {van~der Beek}, \citenamefont
  {Fauqu\'e},\ and\ \citenamefont {Behnia}}]{PhysRevB.92.174504}%
  \BibitemOpen
  \bibfield  {author} {\bibinfo {author} {\bibfnamefont {X.}~\bibnamefont
  {Lin}}, \bibinfo {author} {\bibfnamefont {C.~W.}\ \bibnamefont {Rischau}},
  \bibinfo {author} {\bibfnamefont {C.~J.}\ \bibnamefont {van~der Beek}},
  \bibinfo {author} {\bibfnamefont {B.}~\bibnamefont {Fauqu\'e}},\ and\
  \bibinfo {author} {\bibfnamefont {K.}~\bibnamefont {Behnia}},\ }\href
  {https://doi.org/10.1103/PhysRevB.92.174504} {\bibfield  {journal} {\bibinfo
  {journal} {Phys. Rev. B}\ }\textbf {\bibinfo {volume} {92}},\ \bibinfo
  {pages} {174504} (\bibinfo {year} {2015})}\BibitemShut {NoStop}%
\bibitem [{\citenamefont {Lin}\ \emph {et~al.}(2014{\natexlab{b}})\citenamefont
  {Lin}, \citenamefont {Gourgout}, \citenamefont {Bridoux}, \citenamefont
  {Jomard}, \citenamefont {Pourret}, \citenamefont {Fauqu\'e}, \citenamefont
  {Aoki},\ and\ \citenamefont {Behnia}}]{PhysRevB.90.140508}%
  \BibitemOpen
  \bibfield  {author} {\bibinfo {author} {\bibfnamefont {X.}~\bibnamefont
  {Lin}}, \bibinfo {author} {\bibfnamefont {A.}~\bibnamefont {Gourgout}},
  \bibinfo {author} {\bibfnamefont {G.}~\bibnamefont {Bridoux}}, \bibinfo
  {author} {\bibfnamefont {F.~m.~c.}\ \bibnamefont {Jomard}}, \bibinfo {author}
  {\bibfnamefont {A.}~\bibnamefont {Pourret}}, \bibinfo {author} {\bibfnamefont
  {B.}~\bibnamefont {Fauqu\'e}}, \bibinfo {author} {\bibfnamefont
  {D.}~\bibnamefont {Aoki}},\ and\ \bibinfo {author} {\bibfnamefont
  {K.}~\bibnamefont {Behnia}},\ }\href
  {https://doi.org/10.1103/PhysRevB.90.140508} {\bibfield  {journal} {\bibinfo
  {journal} {Phys. Rev. B}\ }\textbf {\bibinfo {volume} {90}},\ \bibinfo
  {pages} {140508} (\bibinfo {year} {2014}{\natexlab{b}})}\BibitemShut
  {NoStop}%
\bibitem [{\citenamefont {Tinkham}(2004)}]{tinkham2004introduction}%
  \BibitemOpen
  \bibfield  {author} {\bibinfo {author} {\bibfnamefont {M.}~\bibnamefont
  {Tinkham}},\ }\href@noop {} {\emph {\bibinfo {title} {Introduction to
  superconductivity}}}\ (\bibinfo  {publisher} {Dover Publications},\ \bibinfo
  {address} {Mineola, N.Y},\ \bibinfo {year} {2004})\BibitemShut {NoStop}%
\bibitem [{\citenamefont {Marsiglio}(2020)}]{Marsiglio2020}%
  \BibitemOpen
  \bibfield  {author} {\bibinfo {author} {\bibfnamefont {F.}~\bibnamefont
  {Marsiglio}},\ }\href {https://doi.org/10.1016/j.aop.2020.168102} {\bibfield
  {journal} {\bibinfo  {journal} {Annals of Physics}\ }\textbf {\bibinfo
  {volume} {417}},\ \bibinfo {pages} {168102} (\bibinfo {year}
  {2020})}\BibitemShut {NoStop}%
\bibitem [{\citenamefont {Takada}(1980)}]{doi:10.1143/JPSJ.49.1267}%
  \BibitemOpen
  \bibfield  {author} {\bibinfo {author} {\bibfnamefont {Y.}~\bibnamefont
  {Takada}},\ }\href {https://doi.org/10.1143/JPSJ.49.1267} {\bibfield
  {journal} {\bibinfo  {journal} {Journal of the Physical Society of Japan}\
  }\textbf {\bibinfo {volume} {49}},\ \bibinfo {pages} {1267} (\bibinfo {year}
  {1980})},\ \Eprint
  {https://arxiv.org/abs/https://doi.org/10.1143/JPSJ.49.1267}
  {https://doi.org/10.1143/JPSJ.49.1267} \BibitemShut {NoStop}%
\bibitem [{\citenamefont {Kirzhnits}\ \emph {et~al.}(1973)\citenamefont
  {Kirzhnits}, \citenamefont {Maksimov},\ and\ \citenamefont
  {Khomskii}}]{Kirzhnits1973}%
  \BibitemOpen
  \bibfield  {author} {\bibinfo {author} {\bibfnamefont {D.~A.}\ \bibnamefont
  {Kirzhnits}}, \bibinfo {author} {\bibfnamefont {E.~G.}\ \bibnamefont
  {Maksimov}},\ and\ \bibinfo {author} {\bibfnamefont {D.~I.}\ \bibnamefont
  {Khomskii}},\ }\href {https://doi.org/10.1007/BF00655243} {\bibfield
  {journal} {\bibinfo  {journal} {Journal of Low Temperature Physics}\ }\textbf
  {\bibinfo {volume} {10}},\ \bibinfo {pages} {79} (\bibinfo {year}
  {1973})}\BibitemShut {NoStop}%
\bibitem [{\citenamefont {Schooley}\ \emph {et~al.}(1964)\citenamefont
  {Schooley}, \citenamefont {Hosler},\ and\ \citenamefont
  {Cohen}}]{PhysRevLett.12.474}%
  \BibitemOpen
  \bibfield  {author} {\bibinfo {author} {\bibfnamefont {J.~F.}\ \bibnamefont
  {Schooley}}, \bibinfo {author} {\bibfnamefont {W.~R.}\ \bibnamefont
  {Hosler}},\ and\ \bibinfo {author} {\bibfnamefont {M.~L.}\ \bibnamefont
  {Cohen}},\ }\href {https://doi.org/10.1103/PhysRevLett.12.474} {\bibfield
  {journal} {\bibinfo  {journal} {Phys. Rev. Lett.}\ }\textbf {\bibinfo
  {volume} {12}},\ \bibinfo {pages} {474} (\bibinfo {year} {1964})}\BibitemShut
  {NoStop}%
\bibitem [{\citenamefont {Koonce}\ \emph {et~al.}(1967)\citenamefont {Koonce},
  \citenamefont {Cohen}, \citenamefont {Schooley}, \citenamefont {Hosler},\
  and\ \citenamefont {Pfeiffer}}]{PhysRev.163.380}%
  \BibitemOpen
  \bibfield  {author} {\bibinfo {author} {\bibfnamefont {C.~S.}\ \bibnamefont
  {Koonce}}, \bibinfo {author} {\bibfnamefont {M.~L.}\ \bibnamefont {Cohen}},
  \bibinfo {author} {\bibfnamefont {J.~F.}\ \bibnamefont {Schooley}}, \bibinfo
  {author} {\bibfnamefont {W.~R.}\ \bibnamefont {Hosler}},\ and\ \bibinfo
  {author} {\bibfnamefont {E.~R.}\ \bibnamefont {Pfeiffer}},\ }\href
  {https://doi.org/10.1103/PhysRev.163.380} {\bibfield  {journal} {\bibinfo
  {journal} {Phys. Rev.}\ }\textbf {\bibinfo {volume} {163}},\ \bibinfo {pages}
  {380} (\bibinfo {year} {1967})}\BibitemShut {NoStop}%
\bibitem [{\citenamefont {Binnig}\ \emph {et~al.}(1980)\citenamefont {Binnig},
  \citenamefont {Baratoff}, \citenamefont {Hoenig},\ and\ \citenamefont
  {Bednorz}}]{PhysRevLett.45.1352}%
  \BibitemOpen
  \bibfield  {author} {\bibinfo {author} {\bibfnamefont {G.}~\bibnamefont
  {Binnig}}, \bibinfo {author} {\bibfnamefont {A.}~\bibnamefont {Baratoff}},
  \bibinfo {author} {\bibfnamefont {H.~E.}\ \bibnamefont {Hoenig}},\ and\
  \bibinfo {author} {\bibfnamefont {J.~G.}\ \bibnamefont {Bednorz}},\ }\href
  {https://doi.org/10.1103/PhysRevLett.45.1352} {\bibfield  {journal} {\bibinfo
   {journal} {Phys. Rev. Lett.}\ }\textbf {\bibinfo {volume} {45}},\ \bibinfo
  {pages} {1352} (\bibinfo {year} {1980})}\BibitemShut {NoStop}%
\bibitem [{\citenamefont {Gurevich}\ \emph {et~al.}(1962)\citenamefont
  {Gurevich}, \citenamefont {Larkin},\ and\ \citenamefont
  {Firsov}}]{osti_5140639}%
  \BibitemOpen
  \bibfield  {author} {\bibinfo {author} {\bibfnamefont {V.~L.}\ \bibnamefont
  {Gurevich}}, \bibinfo {author} {\bibfnamefont {A.~I.}\ \bibnamefont
  {Larkin}},\ and\ \bibinfo {author} {\bibfnamefont {Y.~A.}\ \bibnamefont
  {Firsov}},\ }\href {https://www.osti.gov/biblio/5140639} {\bibfield
  {journal} {\bibinfo  {journal} {Sov. Phys. - Solid State (Engl. Transl.);
  (United States)}\ }\textbf {\bibinfo {volume} {4}} (\bibinfo {year}
  {1962})}\BibitemShut {NoStop}%
\bibitem [{\citenamefont {Edge}\ \emph {et~al.}(2015)\citenamefont {Edge},
  \citenamefont {Kedem}, \citenamefont {Aschauer}, \citenamefont {Spaldin},\
  and\ \citenamefont {Balatsky}}]{PhysRevLett.115.247002}%
  \BibitemOpen
  \bibfield  {author} {\bibinfo {author} {\bibfnamefont {J.~M.}\ \bibnamefont
  {Edge}}, \bibinfo {author} {\bibfnamefont {Y.}~\bibnamefont {Kedem}},
  \bibinfo {author} {\bibfnamefont {U.}~\bibnamefont {Aschauer}}, \bibinfo
  {author} {\bibfnamefont {N.~A.}\ \bibnamefont {Spaldin}},\ and\ \bibinfo
  {author} {\bibfnamefont {A.~V.}\ \bibnamefont {Balatsky}},\ }\href
  {https://doi.org/10.1103/PhysRevLett.115.247002} {\bibfield  {journal}
  {\bibinfo  {journal} {Phys. Rev. Lett.}\ }\textbf {\bibinfo {volume} {115}},\
  \bibinfo {pages} {247002} (\bibinfo {year} {2015})}\BibitemShut {NoStop}%
\bibitem [{\citenamefont {Fleury}\ \emph {et~al.}(1968)\citenamefont {Fleury},
  \citenamefont {Scott},\ and\ \citenamefont {Worlock}}]{PhysRevLett.21.16}%
  \BibitemOpen
  \bibfield  {author} {\bibinfo {author} {\bibfnamefont {P.~A.}\ \bibnamefont
  {Fleury}}, \bibinfo {author} {\bibfnamefont {J.~F.}\ \bibnamefont {Scott}},\
  and\ \bibinfo {author} {\bibfnamefont {J.~M.}\ \bibnamefont {Worlock}},\
  }\href {https://doi.org/10.1103/PhysRevLett.21.16} {\bibfield  {journal}
  {\bibinfo  {journal} {Phys. Rev. Lett.}\ }\textbf {\bibinfo {volume} {21}},\
  \bibinfo {pages} {16} (\bibinfo {year} {1968})}\BibitemShut {NoStop}%
\bibitem [{\citenamefont {Rischau}\ \emph {et~al.}(2017)\citenamefont
  {Rischau}, \citenamefont {Lin}, \citenamefont {Grams}, \citenamefont {Finck},
  \citenamefont {Harms}, \citenamefont {Engelmayer}, \citenamefont {Lorenz},
  \citenamefont {Gallais}, \citenamefont {Fauqu{\'{e}}}, \citenamefont
  {Hemberger},\ and\ \citenamefont {Behnia}}]{Rischau2017}%
  \BibitemOpen
  \bibfield  {author} {\bibinfo {author} {\bibfnamefont {C.~W.}\ \bibnamefont
  {Rischau}}, \bibinfo {author} {\bibfnamefont {X.}~\bibnamefont {Lin}},
  \bibinfo {author} {\bibfnamefont {C.~P.}\ \bibnamefont {Grams}}, \bibinfo
  {author} {\bibfnamefont {D.}~\bibnamefont {Finck}}, \bibinfo {author}
  {\bibfnamefont {S.}~\bibnamefont {Harms}}, \bibinfo {author} {\bibfnamefont
  {J.}~\bibnamefont {Engelmayer}}, \bibinfo {author} {\bibfnamefont
  {T.}~\bibnamefont {Lorenz}}, \bibinfo {author} {\bibfnamefont
  {Y.}~\bibnamefont {Gallais}}, \bibinfo {author} {\bibfnamefont
  {B.}~\bibnamefont {Fauqu{\'{e}}}}, \bibinfo {author} {\bibfnamefont
  {J.}~\bibnamefont {Hemberger}},\ and\ \bibinfo {author} {\bibfnamefont
  {K.}~\bibnamefont {Behnia}},\ }\href {https://doi.org/10.1038/nphys4085}
  {\bibfield  {journal} {\bibinfo  {journal} {Nature Physics}\ }\textbf
  {\bibinfo {volume} {13}},\ \bibinfo {pages} {643} (\bibinfo {year}
  {2017})}\BibitemShut {NoStop}%
\bibitem [{\citenamefont {Rowley}\ \emph {et~al.}(2014)\citenamefont {Rowley},
  \citenamefont {Spalek}, \citenamefont {Smith}, \citenamefont {Dean},
  \citenamefont {Itoh}, \citenamefont {Scott}, \citenamefont {Lonzarich},\ and\
  \citenamefont {Saxena}}]{Rowley2014}%
  \BibitemOpen
  \bibfield  {author} {\bibinfo {author} {\bibfnamefont {S.~E.}\ \bibnamefont
  {Rowley}}, \bibinfo {author} {\bibfnamefont {L.~J.}\ \bibnamefont {Spalek}},
  \bibinfo {author} {\bibfnamefont {R.~P.}\ \bibnamefont {Smith}}, \bibinfo
  {author} {\bibfnamefont {M.~P.~M.}\ \bibnamefont {Dean}}, \bibinfo {author}
  {\bibfnamefont {M.}~\bibnamefont {Itoh}}, \bibinfo {author} {\bibfnamefont
  {J.~F.}\ \bibnamefont {Scott}}, \bibinfo {author} {\bibfnamefont {G.~G.}\
  \bibnamefont {Lonzarich}},\ and\ \bibinfo {author} {\bibfnamefont {S.~S.}\
  \bibnamefont {Saxena}},\ }\href {https://doi.org/10.1038/nphys2924}
  {\bibfield  {journal} {\bibinfo  {journal} {Nature Physics}\ }\textbf
  {\bibinfo {volume} {10}},\ \bibinfo {pages} {367} (\bibinfo {year}
  {2014})}\BibitemShut {NoStop}%
\bibitem [{\citenamefont {Ahadi}\ \emph {et~al.}(2019)\citenamefont {Ahadi},
  \citenamefont {Galletti}, \citenamefont {Li}, \citenamefont {Salmani-Rezaie},
  \citenamefont {Wu},\ and\ \citenamefont {Stemmer}}]{Ahadi2019}%
  \BibitemOpen
  \bibfield  {author} {\bibinfo {author} {\bibfnamefont {K.}~\bibnamefont
  {Ahadi}}, \bibinfo {author} {\bibfnamefont {L.}~\bibnamefont {Galletti}},
  \bibinfo {author} {\bibfnamefont {Y.}~\bibnamefont {Li}}, \bibinfo {author}
  {\bibfnamefont {S.}~\bibnamefont {Salmani-Rezaie}}, \bibinfo {author}
  {\bibfnamefont {W.}~\bibnamefont {Wu}},\ and\ \bibinfo {author}
  {\bibfnamefont {S.}~\bibnamefont {Stemmer}},\ }\bibfield  {journal} {\bibinfo
   {journal} {Science Advances}\ }\textbf {\bibinfo {volume} {5}},\ \href
  {https://doi.org/10.1126/sciadv.aaw0120} {10.1126/sciadv.aaw0120} (\bibinfo
  {year} {2019})\BibitemShut {NoStop}%
\bibitem [{\citenamefont {Rischau}\ \emph {et~al.}(2022)\citenamefont
  {Rischau}, \citenamefont {Pulmannov\'a}, \citenamefont {Scheerer},
  \citenamefont {Stucky}, \citenamefont {Giannini},\ and\ \citenamefont
  {van~der Marel}}]{PhysRevResearch.4.013019}%
  \BibitemOpen
  \bibfield  {author} {\bibinfo {author} {\bibfnamefont {C.~W.}\ \bibnamefont
  {Rischau}}, \bibinfo {author} {\bibfnamefont {D.}~\bibnamefont
  {Pulmannov\'a}}, \bibinfo {author} {\bibfnamefont {G.~W.}\ \bibnamefont
  {Scheerer}}, \bibinfo {author} {\bibfnamefont {A.}~\bibnamefont {Stucky}},
  \bibinfo {author} {\bibfnamefont {E.}~\bibnamefont {Giannini}},\ and\
  \bibinfo {author} {\bibfnamefont {D.}~\bibnamefont {van~der Marel}},\ }\href
  {https://doi.org/10.1103/PhysRevResearch.4.013019} {\bibfield  {journal}
  {\bibinfo  {journal} {Phys. Rev. Research}\ }\textbf {\bibinfo {volume}
  {4}},\ \bibinfo {pages} {013019} (\bibinfo {year} {2022})}\BibitemShut
  {NoStop}%
\bibitem [{\citenamefont {Mattheiss}(1972{\natexlab{a}})}]{PhysRevB.6.4740}%
  \BibitemOpen
  \bibfield  {author} {\bibinfo {author} {\bibfnamefont {L.~F.}\ \bibnamefont
  {Mattheiss}},\ }\href {https://doi.org/10.1103/PhysRevB.6.4740} {\bibfield
  {journal} {\bibinfo  {journal} {Phys. Rev. B}\ }\textbf {\bibinfo {volume}
  {6}},\ \bibinfo {pages} {4740} (\bibinfo {year}
  {1972}{\natexlab{a}})}\BibitemShut {NoStop}%
\bibitem [{\citenamefont {Guo}\ \emph {et~al.}(2003)\citenamefont {Guo},
  \citenamefont {Chen}, \citenamefont {Sun}, \citenamefont {Sun}, \citenamefont
  {Zhou},\ and\ \citenamefont {Lu}}]{Guo2003}%
  \BibitemOpen
  \bibfield  {author} {\bibinfo {author} {\bibfnamefont {X.}~\bibnamefont
  {Guo}}, \bibinfo {author} {\bibfnamefont {X.}~\bibnamefont {Chen}}, \bibinfo
  {author} {\bibfnamefont {Y.}~\bibnamefont {Sun}}, \bibinfo {author}
  {\bibfnamefont {L.}~\bibnamefont {Sun}}, \bibinfo {author} {\bibfnamefont
  {X.}~\bibnamefont {Zhou}},\ and\ \bibinfo {author} {\bibfnamefont
  {W.}~\bibnamefont {Lu}},\ }\href
  {https://doi.org/10.1016/j.physleta.2003.09.014} {\bibfield  {journal}
  {\bibinfo  {journal} {Physics Letters A}\ }\textbf {\bibinfo {volume}
  {317}},\ \bibinfo {pages} {501} (\bibinfo {year} {2003})}\BibitemShut
  {NoStop}%
\bibitem [{\citenamefont {Piskunov}\ \emph {et~al.}(2004)\citenamefont
  {Piskunov}, \citenamefont {Heifets}, \citenamefont {Eglitis},\ and\
  \citenamefont {Borstel}}]{Piskunov2004}%
  \BibitemOpen
  \bibfield  {author} {\bibinfo {author} {\bibfnamefont {S.}~\bibnamefont
  {Piskunov}}, \bibinfo {author} {\bibfnamefont {E.}~\bibnamefont {Heifets}},
  \bibinfo {author} {\bibfnamefont {R.}~\bibnamefont {Eglitis}},\ and\ \bibinfo
  {author} {\bibfnamefont {G.}~\bibnamefont {Borstel}},\ }\href
  {https://doi.org/10.1016/j.commatsci.2003.08.036} {\bibfield  {journal}
  {\bibinfo  {journal} {Computational Materials Science}\ }\textbf {\bibinfo
  {volume} {29}},\ \bibinfo {pages} {165} (\bibinfo {year} {2004})}\BibitemShut
  {NoStop}%
\bibitem [{\citenamefont {van~der Marel}\ \emph {et~al.}(2011)\citenamefont
  {van~der Marel}, \citenamefont {van Mechelen},\ and\ \citenamefont
  {Mazin}}]{PhysRevB.84.205111}%
  \BibitemOpen
  \bibfield  {author} {\bibinfo {author} {\bibfnamefont {D.}~\bibnamefont
  {van~der Marel}}, \bibinfo {author} {\bibfnamefont {J.~L.~M.}\ \bibnamefont
  {van Mechelen}},\ and\ \bibinfo {author} {\bibfnamefont {I.~I.}\ \bibnamefont
  {Mazin}},\ }\href {https://doi.org/10.1103/PhysRevB.84.205111} {\bibfield
  {journal} {\bibinfo  {journal} {Phys. Rev. B}\ }\textbf {\bibinfo {volume}
  {84}},\ \bibinfo {pages} {205111} (\bibinfo {year} {2011})}\BibitemShut
  {NoStop}%
\bibitem [{\citenamefont {Tao}\ \emph {et~al.}(2016)\citenamefont {Tao},
  \citenamefont {Loret}, \citenamefont {Xu}, \citenamefont {Yang},
  \citenamefont {Rischau}, \citenamefont {Lin}, \citenamefont {Fauqu\'e},
  \citenamefont {Verstraete},\ and\ \citenamefont
  {Behnia}}]{PhysRevB.94.035111}%
  \BibitemOpen
  \bibfield  {author} {\bibinfo {author} {\bibfnamefont {Q.}~\bibnamefont
  {Tao}}, \bibinfo {author} {\bibfnamefont {B.}~\bibnamefont {Loret}}, \bibinfo
  {author} {\bibfnamefont {B.}~\bibnamefont {Xu}}, \bibinfo {author}
  {\bibfnamefont {X.}~\bibnamefont {Yang}}, \bibinfo {author} {\bibfnamefont
  {C.~W.}\ \bibnamefont {Rischau}}, \bibinfo {author} {\bibfnamefont
  {X.}~\bibnamefont {Lin}}, \bibinfo {author} {\bibfnamefont {B.}~\bibnamefont
  {Fauqu\'e}}, \bibinfo {author} {\bibfnamefont {M.~J.}\ \bibnamefont
  {Verstraete}},\ and\ \bibinfo {author} {\bibfnamefont {K.}~\bibnamefont
  {Behnia}},\ }\href {https://doi.org/10.1103/PhysRevB.94.035111} {\bibfield
  {journal} {\bibinfo  {journal} {Phys. Rev. B}\ }\textbf {\bibinfo {volume}
  {94}},\ \bibinfo {pages} {035111} (\bibinfo {year} {2016})}\BibitemShut
  {NoStop}%
\bibitem [{\citenamefont {Shirane}\ and\ \citenamefont
  {Yamada}(1969)}]{PhysRev.177.858}%
  \BibitemOpen
  \bibfield  {author} {\bibinfo {author} {\bibfnamefont {G.}~\bibnamefont
  {Shirane}}\ and\ \bibinfo {author} {\bibfnamefont {Y.}~\bibnamefont
  {Yamada}},\ }\href {https://doi.org/10.1103/PhysRev.177.858} {\bibfield
  {journal} {\bibinfo  {journal} {Phys. Rev.}\ }\textbf {\bibinfo {volume}
  {177}},\ \bibinfo {pages} {858} (\bibinfo {year} {1969})}\BibitemShut
  {NoStop}%
\bibitem [{\citenamefont {Pavarini}\ \emph {et~al.}(2016)\citenamefont
  {Pavarini}, \citenamefont {Koch}, \citenamefont {van~den Brink},\ and\
  \citenamefont {Sawatzky}}]{pavarini2016quantum}%
  \BibitemOpen
  \bibinfo {editor} {\bibfnamefont {E.}~\bibnamefont {Pavarini}}, \bibinfo
  {editor} {\bibfnamefont {E.}~\bibnamefont {Koch}}, \bibinfo {editor}
  {\bibfnamefont {J.}~\bibnamefont {van~den Brink}},\ and\ \bibinfo {editor}
  {\bibfnamefont {G.}~\bibnamefont {Sawatzky}},\ eds.,\ \href@noop {} {\emph
  {\bibinfo {title} {Quantum materials: Experiments and theory: Lecture notes
  of the autumn school on correlated electrons at Forschungszentrum Julich}}}\
  (\bibinfo {year} {2016})\BibitemShut {NoStop}%
\bibitem [{\citenamefont {Harrison}(1989)}]{Harrison}%
  \BibitemOpen
  \bibfield  {author} {\bibinfo {author} {\bibfnamefont {W.~A.}\ \bibnamefont
  {Harrison}},\ }\href@noop {} {\emph {\bibinfo {title} {Electronic Structure
  and the Properties of Solids: The Physics of the Chemical Bond}}}\ (\bibinfo
  {publisher} {Dover Publications},\ \bibinfo {address} {New York},\ \bibinfo
  {year} {1989})\BibitemShut {NoStop}%
\bibitem [{\citenamefont {Mattheiss}(1972{\natexlab{b}})}]{PhysRevB.6.4718}%
  \BibitemOpen
  \bibfield  {author} {\bibinfo {author} {\bibfnamefont {L.~F.}\ \bibnamefont
  {Mattheiss}},\ }\href {https://doi.org/10.1103/PhysRevB.6.4718} {\bibfield
  {journal} {\bibinfo  {journal} {Phys. Rev. B}\ }\textbf {\bibinfo {volume}
  {6}},\ \bibinfo {pages} {4718} (\bibinfo {year}
  {1972}{\natexlab{b}})}\BibitemShut {NoStop}%
\bibitem [{\citenamefont {Dunn}(1961)}]{Dunn1961-ce}%
  \BibitemOpen
  \bibfield  {author} {\bibinfo {author} {\bibfnamefont {T.~M.}\ \bibnamefont
  {Dunn}},\ }\href@noop {} {\bibfield  {journal} {\bibinfo  {journal} {Trans.
  Faraday Soc.}\ }\textbf {\bibinfo {volume} {57}},\ \bibinfo {pages} {1441}
  (\bibinfo {year} {1961})}\BibitemShut {NoStop}%
\bibitem [{\citenamefont {Landau}(1977)}]{landau1977quantum}%
  \BibitemOpen
  \bibfield  {author} {\bibinfo {author} {\bibfnamefont {L.~D.}\ \bibnamefont
  {Landau}},\ }\href@noop {} {\emph {\bibinfo {title} {Quantum mechanics:
  non-relativistic theory}}}\ (\bibinfo  {publisher} {Pergamon Press},\
  \bibinfo {address} {Oxford New York},\ \bibinfo {year} {1977})\BibitemShut
  {NoStop}%
\bibitem [{\citenamefont {Dresselhaus}\ \emph {et~al.}(2007)\citenamefont
  {Dresselhaus}, \citenamefont {Dresselhaus},\ and\ \citenamefont
  {Jorio}}]{dresselhaus2007group}%
  \BibitemOpen
  \bibfield  {author} {\bibinfo {author} {\bibfnamefont {M.}~\bibnamefont
  {Dresselhaus}}, \bibinfo {author} {\bibfnamefont {G.}~\bibnamefont
  {Dresselhaus}},\ and\ \bibinfo {author} {\bibfnamefont {A.}~\bibnamefont
  {Jorio}},\ }\href {https://books.google.com/books?id=SPiqKWk\_h2sC} {\emph
  {\bibinfo {title} {Group Theory: Application to the Physics of Condensed
  Matter}}},\ SpringerLink: Springer e-Books\ (\bibinfo  {publisher} {Springer
  Berlin Heidelberg},\ \bibinfo {year} {2007})\BibitemShut {NoStop}%
\bibitem [{\citenamefont {Lin}\ \emph {et~al.}(2013)\citenamefont {Lin},
  \citenamefont {Zhu}, \citenamefont {Fauque},\ and\ \citenamefont
  {Behnia}}]{PhysRevX.3.021002}%
  \BibitemOpen
  \bibfield  {author} {\bibinfo {author} {\bibfnamefont {X.}~\bibnamefont
  {Lin}}, \bibinfo {author} {\bibfnamefont {Z.}~\bibnamefont {Zhu}}, \bibinfo
  {author} {\bibfnamefont {B.}~\bibnamefont {Fauque}},\ and\ \bibinfo {author}
  {\bibfnamefont {K.}~\bibnamefont {Behnia}},\ }\href
  {https://doi.org/10.1103/PhysRevX.3.021002} {\bibfield  {journal} {\bibinfo
  {journal} {Phys. Rev. X}\ }\textbf {\bibinfo {volume} {3}},\ \bibinfo {pages}
  {021002} (\bibinfo {year} {2013})}\BibitemShut {NoStop}%
\bibitem [{\citenamefont {Dunnett}\ \emph {et~al.}(2018)\citenamefont
  {Dunnett}, \citenamefont {Narayan}, \citenamefont {Spaldin},\ and\
  \citenamefont {Balatsky}}]{PhysRevB.97.144506}%
  \BibitemOpen
  \bibfield  {author} {\bibinfo {author} {\bibfnamefont {K.}~\bibnamefont
  {Dunnett}}, \bibinfo {author} {\bibfnamefont {A.}~\bibnamefont {Narayan}},
  \bibinfo {author} {\bibfnamefont {N.~A.}\ \bibnamefont {Spaldin}},\ and\
  \bibinfo {author} {\bibfnamefont {A.~V.}\ \bibnamefont {Balatsky}},\ }\href
  {https://doi.org/10.1103/PhysRevB.97.144506} {\bibfield  {journal} {\bibinfo
  {journal} {Phys. Rev. B}\ }\textbf {\bibinfo {volume} {97}},\ \bibinfo
  {pages} {144506} (\bibinfo {year} {2018})}\BibitemShut {NoStop}%
\bibitem [{\citenamefont {Kedem}(2020)}]{kedem2020paradigm}%
  \BibitemOpen
  \bibfield  {author} {\bibinfo {author} {\bibfnamefont {Y.}~\bibnamefont
  {Kedem}},\ }\href@noop {} {\bibfield  {journal} {\bibinfo  {journal}
  {arXiv:2004.00029}\ } (\bibinfo {year} {2020})}\BibitemShut {NoStop}%
\bibitem [{\citenamefont {Omar}\ \emph {et~al.}(2021)\citenamefont {Omar},
  \citenamefont {Kong}, \citenamefont {Jani}, \citenamefont {Li}, \citenamefont
  {Zhou}, \citenamefont {Lim}, \citenamefont {Prakash}, \citenamefont {Zeng},
  \citenamefont {Hooda}, \citenamefont {Venkatesan}, \citenamefont {Feng},
  \citenamefont {Pennycook}, \citenamefont {Lei},\ and\ \citenamefont
  {Ariando}}]{omar2021large}%
  \BibitemOpen
  \bibfield  {author} {\bibinfo {author} {\bibfnamefont {G.~J.}\ \bibnamefont
  {Omar}}, \bibinfo {author} {\bibfnamefont {W.}~\bibnamefont {Kong}}, \bibinfo
  {author} {\bibfnamefont {H.}~\bibnamefont {Jani}}, \bibinfo {author}
  {\bibfnamefont {M.}~\bibnamefont {Li}}, \bibinfo {author} {\bibfnamefont
  {J.}~\bibnamefont {Zhou}}, \bibinfo {author} {\bibfnamefont {Z.~S.}\
  \bibnamefont {Lim}}, \bibinfo {author} {\bibfnamefont {S.}~\bibnamefont
  {Prakash}}, \bibinfo {author} {\bibfnamefont {S.}~\bibnamefont {Zeng}},
  \bibinfo {author} {\bibfnamefont {S.}~\bibnamefont {Hooda}}, \bibinfo
  {author} {\bibfnamefont {T.}~\bibnamefont {Venkatesan}}, \bibinfo {author}
  {\bibfnamefont {Y.~P.}\ \bibnamefont {Feng}}, \bibinfo {author}
  {\bibfnamefont {S.~J.}\ \bibnamefont {Pennycook}}, \bibinfo {author}
  {\bibfnamefont {S.}~\bibnamefont {Lei}},\ and\ \bibinfo {author}
  {\bibfnamefont {A.}~\bibnamefont {Ariando}},\ }\href@noop {} {\bibfield
  {journal} {\bibinfo  {journal} {arXiv:2110.06728}\ } (\bibinfo {year}
  {2021})}\BibitemShut {NoStop}%
\bibitem [{\citenamefont {Schumann}\ \emph {et~al.}(2020)\citenamefont
  {Schumann}, \citenamefont {Galletti}, \citenamefont {Jeong}, \citenamefont
  {Ahadi}, \citenamefont {Strickland}, \citenamefont {Salmani-Rezaie},\ and\
  \citenamefont {Stemmer}}]{PhysRevB.101.100503}%
  \BibitemOpen
  \bibfield  {author} {\bibinfo {author} {\bibfnamefont {T.}~\bibnamefont
  {Schumann}}, \bibinfo {author} {\bibfnamefont {L.}~\bibnamefont {Galletti}},
  \bibinfo {author} {\bibfnamefont {H.}~\bibnamefont {Jeong}}, \bibinfo
  {author} {\bibfnamefont {K.}~\bibnamefont {Ahadi}}, \bibinfo {author}
  {\bibfnamefont {W.~M.}\ \bibnamefont {Strickland}}, \bibinfo {author}
  {\bibfnamefont {S.}~\bibnamefont {Salmani-Rezaie}},\ and\ \bibinfo {author}
  {\bibfnamefont {S.}~\bibnamefont {Stemmer}},\ }\href
  {https://doi.org/10.1103/PhysRevB.101.100503} {\bibfield  {journal} {\bibinfo
   {journal} {Phys. Rev. B}\ }\textbf {\bibinfo {volume} {101}},\ \bibinfo
  {pages} {100503} (\bibinfo {year} {2020})}\BibitemShut {NoStop}%
\bibitem [{\citenamefont {Mott}(1949)}]{Mott1949-pz}%
  \BibitemOpen
  \bibfield  {author} {\bibinfo {author} {\bibfnamefont {N.~F.}\ \bibnamefont
  {Mott}},\ }\href@noop {} {\bibfield  {journal} {\bibinfo  {journal} {Proc.
  Phys. Soc.}\ }\textbf {\bibinfo {volume} {62}},\ \bibinfo {pages} {416}
  (\bibinfo {year} {1949})}\BibitemShut {NoStop}%
\bibitem [{\citenamefont {Anderson}(1987)}]{doi:10.1126/science.235.4793.1196}%
  \BibitemOpen
  \bibfield  {author} {\bibinfo {author} {\bibfnamefont {P.~W.}\ \bibnamefont
  {Anderson}},\ }\href {https://doi.org/10.1126/science.235.4793.1196}
  {\bibfield  {journal} {\bibinfo  {journal} {Science}\ }\textbf {\bibinfo
  {volume} {235}},\ \bibinfo {pages} {1196} (\bibinfo {year}
  {1987})}\BibitemShut {NoStop}%
\bibitem [{\citenamefont {de’ Medici}\ and\ \citenamefont
  {Capone}(2017)}]{de_medici_2017}%
  \BibitemOpen
  \bibfield  {author} {\bibinfo {author} {\bibfnamefont {L.}~\bibnamefont
  {de’ Medici}}\ and\ \bibinfo {author} {\bibfnamefont {M.}~\bibnamefont
  {Capone}},\ }\href {https://doi.org/10.1007/978-3-319-56117-2_4} {\bibfield
  {journal} {\bibinfo  {journal} {Springer Series in Solid-State Sciences}\ ,\
  \bibinfo {pages} {115–185}} (\bibinfo {year} {2017})}\BibitemShut {NoStop}%
\bibitem [{\citenamefont {de' Medici}(2011)}]{PhysRevB.83.205112}%
  \BibitemOpen
  \bibfield  {author} {\bibinfo {author} {\bibfnamefont {L.}~\bibnamefont {de'
  Medici}},\ }\href {https://doi.org/10.1103/PhysRevB.83.205112} {\bibfield
  {journal} {\bibinfo  {journal} {Phys. Rev. B}\ }\textbf {\bibinfo {volume}
  {83}},\ \bibinfo {pages} {205112} (\bibinfo {year} {2011})}\BibitemShut
  {NoStop}%
\bibitem [{\citenamefont {Allmaier}\ \emph {et~al.}(2009)\citenamefont
  {Allmaier}, \citenamefont {Chioncel},\ and\ \citenamefont
  {Arrigoni}}]{PhysRevB.79.235126}%
  \BibitemOpen
  \bibfield  {author} {\bibinfo {author} {\bibfnamefont {H.}~\bibnamefont
  {Allmaier}}, \bibinfo {author} {\bibfnamefont {L.}~\bibnamefont {Chioncel}},\
  and\ \bibinfo {author} {\bibfnamefont {E.}~\bibnamefont {Arrigoni}},\ }\href
  {https://doi.org/10.1103/PhysRevB.79.235126} {\bibfield  {journal} {\bibinfo
  {journal} {Phys. Rev. B}\ }\textbf {\bibinfo {volume} {79}},\ \bibinfo
  {pages} {235126} (\bibinfo {year} {2009})}\BibitemShut {NoStop}%
\bibitem [{\citenamefont {MOTT}(1968)}]{RevModPhys.40.677}%
  \BibitemOpen
  \bibfield  {author} {\bibinfo {author} {\bibfnamefont {N.~F.}\ \bibnamefont
  {MOTT}},\ }\href {https://doi.org/10.1103/RevModPhys.40.677} {\bibfield
  {journal} {\bibinfo  {journal} {Rev. Mod. Phys.}\ }\textbf {\bibinfo {volume}
  {40}},\ \bibinfo {pages} {677} (\bibinfo {year} {1968})}\BibitemShut
  {NoStop}%
\bibitem [{\citenamefont {Edwards}\ and\ \citenamefont
  {Sienko}(1978)}]{PhysRevB.17.2575}%
  \BibitemOpen
  \bibfield  {author} {\bibinfo {author} {\bibfnamefont {P.~P.}\ \bibnamefont
  {Edwards}}\ and\ \bibinfo {author} {\bibfnamefont {M.~J.}\ \bibnamefont
  {Sienko}},\ }\href {https://doi.org/10.1103/PhysRevB.17.2575} {\bibfield
  {journal} {\bibinfo  {journal} {Phys. Rev. B}\ }\textbf {\bibinfo {volume}
  {17}},\ \bibinfo {pages} {2575} (\bibinfo {year} {1978})}\BibitemShut
  {NoStop}%
\bibitem [{not()}]{note}%
  \BibitemOpen
  \href@noop {} {}\bibinfo {note} {See Appendix for an alternative derivation
  that does not rely on an ad-hoc wavefunction}\BibitemShut {NoStop}%
\bibitem [{\citenamefont {King}\ \emph {et~al.}(2014)\citenamefont {King},
  \citenamefont {Walker}, \citenamefont {Tamai}, \citenamefont {de~la Torre},
  \citenamefont {Eknapakul}, \citenamefont {Buaphet}, \citenamefont {Mo},
  \citenamefont {Meevasana}, \citenamefont {Bahramy},\ and\ \citenamefont
  {Baumberger}}]{King2014}%
  \BibitemOpen
  \bibfield  {author} {\bibinfo {author} {\bibfnamefont {P.~D.~C.}\
  \bibnamefont {King}}, \bibinfo {author} {\bibfnamefont {S.~M.}\ \bibnamefont
  {Walker}}, \bibinfo {author} {\bibfnamefont {A.}~\bibnamefont {Tamai}},
  \bibinfo {author} {\bibfnamefont {A.}~\bibnamefont {de~la Torre}}, \bibinfo
  {author} {\bibfnamefont {T.}~\bibnamefont {Eknapakul}}, \bibinfo {author}
  {\bibfnamefont {P.}~\bibnamefont {Buaphet}}, \bibinfo {author} {\bibfnamefont
  {S.-K.}\ \bibnamefont {Mo}}, \bibinfo {author} {\bibfnamefont
  {W.}~\bibnamefont {Meevasana}}, \bibinfo {author} {\bibfnamefont {M.~S.}\
  \bibnamefont {Bahramy}},\ and\ \bibinfo {author} {\bibfnamefont
  {F.}~\bibnamefont {Baumberger}},\ }\bibfield  {journal} {\bibinfo  {journal}
  {Nature Communications}\ }\textbf {\bibinfo {volume} {5}},\ \href
  {https://doi.org/10.1038/ncomms4414} {10.1038/ncomms4414} (\bibinfo {year}
  {2014})\BibitemShut {NoStop}%
\bibitem [{\citenamefont {Walker}\ \emph {et~al.}(2015)\citenamefont {Walker},
  \citenamefont {Bruno}, \citenamefont {Wang}, \citenamefont {de~la Torre},
  \citenamefont {Ricc{\'{o}}}, \citenamefont {Tamai}, \citenamefont {Kim},
  \citenamefont {Hoesch}, \citenamefont {Shi}, \citenamefont {Bahramy},
  \citenamefont {King},\ and\ \citenamefont {Baumberger}}]{Walker2015}%
  \BibitemOpen
  \bibfield  {author} {\bibinfo {author} {\bibfnamefont {S.~M.}\ \bibnamefont
  {Walker}}, \bibinfo {author} {\bibfnamefont {F.~Y.}\ \bibnamefont {Bruno}},
  \bibinfo {author} {\bibfnamefont {Z.}~\bibnamefont {Wang}}, \bibinfo {author}
  {\bibfnamefont {A.}~\bibnamefont {de~la Torre}}, \bibinfo {author}
  {\bibfnamefont {S.}~\bibnamefont {Ricc{\'{o}}}}, \bibinfo {author}
  {\bibfnamefont {A.}~\bibnamefont {Tamai}}, \bibinfo {author} {\bibfnamefont
  {T.~K.}\ \bibnamefont {Kim}}, \bibinfo {author} {\bibfnamefont
  {M.}~\bibnamefont {Hoesch}}, \bibinfo {author} {\bibfnamefont
  {M.}~\bibnamefont {Shi}}, \bibinfo {author} {\bibfnamefont {M.~S.}\
  \bibnamefont {Bahramy}}, \bibinfo {author} {\bibfnamefont {P.~D.~C.}\
  \bibnamefont {King}},\ and\ \bibinfo {author} {\bibfnamefont
  {F.}~\bibnamefont {Baumberger}},\ }\href
  {https://doi.org/10.1002/adma.201501556} {\bibfield  {journal} {\bibinfo
  {journal} {Advanced Materials}\ }\textbf {\bibinfo {volume} {27}},\ \bibinfo
  {pages} {3894} (\bibinfo {year} {2015})}\BibitemShut {NoStop}%
\bibitem [{\citenamefont {Balents}\ \emph {et~al.}(2020)\citenamefont
  {Balents}, \citenamefont {Dean}, \citenamefont {Efetov},\ and\ \citenamefont
  {Young}}]{Balents2020-gp}%
  \BibitemOpen
  \bibfield  {author} {\bibinfo {author} {\bibfnamefont {L.}~\bibnamefont
  {Balents}}, \bibinfo {author} {\bibfnamefont {C.~R.}\ \bibnamefont {Dean}},
  \bibinfo {author} {\bibfnamefont {D.~K.}\ \bibnamefont {Efetov}},\ and\
  \bibinfo {author} {\bibfnamefont {A.~F.}\ \bibnamefont {Young}},\ }\href@noop
  {} {\bibfield  {journal} {\bibinfo  {journal} {Nat. Phys.}\ }\textbf
  {\bibinfo {volume} {16}},\ \bibinfo {pages} {725} (\bibinfo {year}
  {2020})}\BibitemShut {NoStop}%
\bibitem [{\citenamefont {Cooper}(1956)}]{PhysRev.104.1189}%
  \BibitemOpen
  \bibfield  {author} {\bibinfo {author} {\bibfnamefont {L.~N.}\ \bibnamefont
  {Cooper}},\ }\href {https://doi.org/10.1103/PhysRev.104.1189} {\bibfield
  {journal} {\bibinfo  {journal} {Phys. Rev.}\ }\textbf {\bibinfo {volume}
  {104}},\ \bibinfo {pages} {1189} (\bibinfo {year} {1956})}\BibitemShut
  {NoStop}%
\bibitem [{\citenamefont {Collignon}\ \emph {et~al.}(2020)\citenamefont
  {Collignon}, \citenamefont {Bourges}, \citenamefont {Fauqu\'e},\ and\
  \citenamefont {Behnia}}]{PhysRevX.10.031025}%
  \BibitemOpen
  \bibfield  {author} {\bibinfo {author} {\bibfnamefont {C.}~\bibnamefont
  {Collignon}}, \bibinfo {author} {\bibfnamefont {P.}~\bibnamefont {Bourges}},
  \bibinfo {author} {\bibfnamefont {B.}~\bibnamefont {Fauqu\'e}},\ and\
  \bibinfo {author} {\bibfnamefont {K.}~\bibnamefont {Behnia}},\ }\href
  {https://doi.org/10.1103/PhysRevX.10.031025} {\bibfield  {journal} {\bibinfo
  {journal} {Phys. Rev. X}\ }\textbf {\bibinfo {volume} {10}},\ \bibinfo
  {pages} {031025} (\bibinfo {year} {2020})}\BibitemShut {NoStop}%
\bibitem [{\citenamefont {Ohtomo}\ and\ \citenamefont
  {Hwang}(2004)}]{Ohtomo2004}%
  \BibitemOpen
  \bibfield  {author} {\bibinfo {author} {\bibfnamefont {A.}~\bibnamefont
  {Ohtomo}}\ and\ \bibinfo {author} {\bibfnamefont {H.~Y.}\ \bibnamefont
  {Hwang}},\ }\href {https://doi.org/10.1038/nature02308} {\bibfield  {journal}
  {\bibinfo  {journal} {Nature}\ }\textbf {\bibinfo {volume} {427}},\ \bibinfo
  {pages} {423} (\bibinfo {year} {2004})}\BibitemShut {NoStop}%
\bibitem [{\citenamefont {Glinchuk}\ \emph {et~al.}(2001)\citenamefont
  {Glinchuk}, \citenamefont {Bykov}, \citenamefont {Slipenyuk}, \citenamefont
  {Laguta},\ and\ \citenamefont {Jastrabik}}]{Glinchuk2001}%
  \BibitemOpen
  \bibfield  {author} {\bibinfo {author} {\bibfnamefont {M.~D.}\ \bibnamefont
  {Glinchuk}}, \bibinfo {author} {\bibfnamefont {I.~P.}\ \bibnamefont {Bykov}},
  \bibinfo {author} {\bibfnamefont {A.~M.}\ \bibnamefont {Slipenyuk}}, \bibinfo
  {author} {\bibfnamefont {V.~V.}\ \bibnamefont {Laguta}},\ and\ \bibinfo
  {author} {\bibfnamefont {L.}~\bibnamefont {Jastrabik}},\ }\href
  {https://doi.org/10.1134/1.1371362} {\bibfield  {journal} {\bibinfo
  {journal} {Physics of the Solid State}\ }\textbf {\bibinfo {volume} {43}},\
  \bibinfo {pages} {841} (\bibinfo {year} {2001})}\BibitemShut {NoStop}%
\bibitem [{\citenamefont {Baskaran}\ \emph {et~al.}(1987)\citenamefont
  {Baskaran}, \citenamefont {Zou},\ and\ \citenamefont
  {Anderson}}]{BASKARAN1987973}%
  \BibitemOpen
  \bibfield  {author} {\bibinfo {author} {\bibfnamefont {G.}~\bibnamefont
  {Baskaran}}, \bibinfo {author} {\bibfnamefont {Z.}~\bibnamefont {Zou}},\ and\
  \bibinfo {author} {\bibfnamefont {P.}~\bibnamefont {Anderson}},\ }\href
  {https://doi.org/https://doi.org/10.1016/0038-1098(87)90642-9} {\bibfield
  {journal} {\bibinfo  {journal} {Solid State Communications}\ }\textbf
  {\bibinfo {volume} {63}},\ \bibinfo {pages} {973} (\bibinfo {year}
  {1987})}\BibitemShut {NoStop}%
\end{thebibliography}%
\bibliographystyle{apsrev4-2}
\end{document}